\documentclass[showpacs,floatfix,aps]{revtex4}

\newcommand{\beq}{\begin{equation}}
\newcommand{\eeq}{\end{equation}}
\newcommand{\beqa}{\begin{eqnarray}}
\newcommand{\eeqa}{\end{eqnarray}}
\newcommand{\sla}[1]{\not\!#1}
\newcommand{\tdpi}{\vtau \cdot \vpi}
\newcommand{\vtau}{\vec{\tau}}
\newcommand{\vpi}{\vec{\pi}}
\newcommand{\F}{F_0}
\newcommand{\Fpi}{F_\pi}
\newcommand{\dv}{\partial}
\newcommand{\dlmu}{\partial_\mu}
\newcommand{\nn}{\nonumber}
\newcommand{\eqindent}{\quad \quad \quad}

\newcommand{\tdvv}{\vtau \cdot \vec{v}}
\newcommand{\tdva}{\vtau \cdot \vec{a}}
\newcommand{\Psibn}{\overline{\Psi}_f}
\newcommand{\Psip}{\Psi_i}
\newcommand{\Psib}{\overline{\Psi}}
\newcommand{\lnmpi}{{\rm ln}(\frac{m_\pi^2}{m_N^2})}
\newcommand{\lnmpib}{{\rm ln}(\frac{m_{0\pi}^2}{\mu^2})}
\newcommand{\LogmN}{{\rm ln}(\frac{m_N^2}{\mu^2})}
\newcommand{\Logm}{{\rm ln}(\frac{m^2}{\mu^2})}
\newcommand{\BL}{\beta^{IR}}
\newcommand{\EO}{\beta^{EOMS}}
\newcommand{\FPsq}{F_\pi^2 \pi^2}
\newcommand{\dm}{\delta_m}

\usepackage{epsfig}

\begin{document}

\hfill{TRI-PP-06-11} 

\title{Ordinary muon capture on a proton 
in manifestly Lorentz invariant baryon chiral perturbation theory}

\author{Shung-ichi Ando\footnote{Current address: Department of Physics, 
Sungkyunkwan University, Suwon 440-746, 
Korea}\footnote{E-mail:sando@meson.skku.ac.kr} and 
Harold W. Fearing\footnote{E-mail:fearing@triumf.ca}}

\affiliation{TRIUMF, 
Vancouver, 
British Columbia, V6T 2A3, Canada}

\begin{abstract}
The amplitude for ordinary muon capture on the proton is evaluated,
through the first four orders in the expansion parameter, in a
manifestly Lorentz invariant form of baryon chiral perturbation
theory. Expressions for the low energy constants in terms of physical
quantities are obtained in each of the several renormalization schemes
which have been proposed for forcing the relativistic approach to obey
the same counting rules as obtained in heavy baryon chiral
perturbation theory. The advantages and disadvantages of these schemes
are discussed, using the muon capture results as an example, with the
aim of gaining insight as to which scheme is preferable for practical
calculations.
\end{abstract}


\maketitle

\section{Introduction}
Ordinary, or non radiative, muon capture (OMC) has always been an
interesting process because, unlike beta decay, there is a
sufficiently high momentum transfer to explore weak-nucleon form
factors away from $q^2=0$, where $q^\mu$ is the four momentum transfer
between lepton and hadron currents. The induced pseudoscalar form
factor, $G_P$, has been of particular interest and OMC is the main
source of information on this form factor, see, e.~g.,
Refs.~\cite{Gorrev,bem-jpg02,m-pr01}. Originally the amplitude for OMC
was written down as the most general form of the current and OMC was
simply an empirical way to determine the coefficients of this most
general current. More recently it has become possible to calculate
these form factors from a more fundamental point of view using an
effective Lagrangian of chiral perturbation theory (ChPT) which
incorporates the symmetries of QCD \cite{w-pa79}.  Such calculations
for OMC \cite{fear97,amk-prc01,bhm-npa01} or for the electromagnetic
form factors of the nucleon \cite{bfhm-npa98} have been carried out in
so called heavy baryon ChPT (HBChPT) which involves a
Foldy-Wouthuysen-like expansion of the Lagrangian in powers of the
inverse nucleon mass. A comprehensive, modern review with references to
earlier work can be found in \cite{Srev}.

The possibility of a fully relativistic ChPT approach to OMC, or any
other process, has been elusive until recently. However there now have
been several suggestions \cite{fuchs03,Sch04,BL,et-prc98,geg03} for
relativistic approaches \cite{footnote1}. The difficulty with
relativistic theories has been the fact, as pointed out by Gasser,
{\it et al.} \cite{Gasser}, that they do not obey a counting procedure
which would allow one to associate multiple loop diagrams with higher
powers in an expansion in a small parameter. In fact in the
relativistic approach these multiple loop diagrams, specifically those
involving the relativistic nucleon propagator, can contribute to lower
orders and so there is not a well defined prescription for deciding
what diagrams to keep. One solution to this problem was the HBChPT
approach in which diagrams involving more and more loops contribute at
higher and higher orders in an expansion of the amplitude in powers of
a typical momentum scale divided by the nucleon mass.  While this
approach works, it has the disadvantage of not being manifestly
Lorentz invariant and of requiring increasingly complicated vertices
as the order increases. The recent proposals for relativistic ChPT
resolve this problem in a different way by showing that it is possible
to define renormalization procedures for a manifestly Lorentz
invariant theory which generate the same type of counting scheme which
is present in HBChPT.

This paper thus has a number of aims. OMC is one of the simplest non
trivial processes, and so we want to use it as laboratory to
understand how the relativistic approach and the various
renormalization schemes are applied to a practical case. We also want
to compare and contrast the various proposed schemes to see if one is
preferable for detailed calculations or if there are alternative
methods which achieve the same result but which are easier to use.
 
We also want to obtain the muon capture amplitude and expressions for
the low energy constants (LEC's) in a unified and consistent
approach. In the relativistic approach one can obtain these quantities
to one higher order than is easily possible in HBChPT. This results
from the fact that the Lagrangian, and the vertex operators
originating from it, increases in complexity with increasing order
much more rapidly for HBChPT than for the relativistic approach.

The OMC amplitude accesses the weak nucleon currents, vector and axial
vector, the former being essentially the same as the electromagnetic
current. There is enough information available to determine all of the
LEC's which appear. However, just as in HBChPT, the only unused
information, once the LEC's are determined, serves only to give the
well known expression for $G_P$ in terms of the pion-nucleon coupling
and the axial radius. Thus this evaluation of the OMC amplitude is
mainly a way of determining the LEC's and does not provide a 'new'
number for the rate.

Although many of the pieces have been obtained before in separate
calculations,
\cite{Gasser,BLpiN,fgs-jpg04,Fuchsthesis,km-npa01,Schwthesis} 
the results for the LEC's depend on the details of
the calculation and it becomes dangerous to lift these values for the
LEC's from disparate calculations and use them for other
processes. Therefore it is important to have a consistent,
consolidated and practical approach as this will provide the basis for
determining the LEC's which will be necessary for future
calculations. Thus we see this calculation as a basis or starting
point for planned similar consistent applications of this approach to
processes such as $\pi p \rightarrow n \gamma$ where there are puzzles
in the HBChPT approach arising from the appearance of unnaturally
large LEC's \cite{Fearpirad}. It is also intended to be a starting
point for a similar calculation of radiative muon capture, 
$\mu + p \rightarrow n + \nu + \gamma$,
\cite{mmk-plb98,am-plb98}, where there are still unresolved problems
relative to the extraction of $G_P$
\cite{Gorrev,afm-prc01,tk-prc02,amk-prc02}.

\section{Weak form factors of the nucleon current}
The S-matrix amplitude for the OMC process, $\mu+p\to n+\nu$, with momenta 
defined by $p_\mu+p_i=p_f+p_\nu$, is given in the notation of \cite{BjD} by
\beq \label{OMCamp}
   {\cal M} = \frac{-iG_F V_{ud}}{\sqrt{2}}\overline{u}(p_\nu)\gamma_
       \alpha(1-\gamma_5)u(p_\mu)\overline{u}(p_f)
       \tau_-\left[V^\alpha-A^\alpha\right]u(p_i)
\eeq
where the vector and axial vector nucleon current operators are given by
\beqa
V^\alpha &=&G_V(q^2)\gamma^\alpha
    +\frac{iG_M(q^2)}{2m_N}\sigma^{\alpha\beta}q_\beta \nonumber \, , \\
A^\alpha &=& G_A(q^2)\gamma^\alpha\gamma_5
                  +\frac{G_P(q^2)}{m_\mu}q^\alpha\gamma_5 \, .
\eeqa
Here $G_F$ is the Fermi constant as obtained from muon decay, $V_{ud}$
is an element of the CKM matrix, $m_\mu$ is the physical muon mass,
$m_N$ is the average of physical neutron and proton masses,
$m_N=(m_n+m_p)/2$, and $\tau_-$ is the isospin lowering operator,
$<n|\tau_-|p>=1$.  Here we do not include 
radiative corrections \cite{aetal-plb04} and have neglected possible 
second class currents.

$G_V(q^2)$, $G_M(q^2)$, $G_A(q^2)$ and $G_P(q^2)$ are the weak form
factors of interest, with the four momentum transfer
$q^\mu=p_f^\mu-p_i^\mu$. The vector and weak magnetism form factors,
$G_V$ and $G_M$ respectively, are related to the isovector
electromagnetic form factors of the nucleon. The axial form factor at
$q^2=0$, $G_A(0)$, is most accurately determined from neutron beta
decay and $G_P$ is the induced pseudoscalar form factor which is
accurately predicted by chiral symmetry. All of these form factors are
functions of the four momentum transfer $q^2$ which for OMC on the
proton is given by
\beq 
q^2 \rightarrow \frac{-m_\mu(m_p^2-m_n^2+m_\mu m_p)}{m_p+m_\mu} = -0.88
m_\mu^2.  
\eeq
Note that we have normalized these form factors using the physical
masses $m_\mu$ and $m_N$ and that we have used the ChPT sign
convention for $G_A$ and $G_P$, which makes them positive, in contrast
to the convention which has been used historically and which is still
used in the Particle Data Group listings \cite{PDG}.

\section{Effective chiral Lagrangian}
\subsection{Effective Lagrangian}
In the usual ChPT approach the effective Lagrangian is expanded in
powers of a typical momentum - for this problem the muon mass $m_\mu$,
the pion mass $m_\pi$, or the four momentum transfer squared $q^2$ -
divided by a typical hadronic scale which we take as the physical
nucleon mass. At each order the most general Lagrangian satisfying the
symmetries of QCD is determined.

For this calculation we work in SU(2)$\times$SU(2) and use for the
chiral Lagrangian ${\cal L}_\chi = {\cal L}_{\pi N} + {\cal L}_\pi$
where ${\cal L}_{\pi N}$ and ${\cal L}_{\pi}$ are respectively the
Lagrangians in the pion-nucleon and pion sectors.  The pion-nucleon
Lagrangian is expanded in terms of small quantities,
\beq
{\cal L}_{\pi N} = {\cal L}_{\pi N}^{(1)}+{\cal L}_{\pi N}^{(2)}+
{\cal L}_{\pi N}^{(3)}+{\cal L}_{\pi N}^{(4)}+\cdots \, ,
\eeq
where the ellipsis represents the higher order terms and  
the superscript denotes the order of the Lagrangian.

The lowest order Lagrangian is given by the standard form
\beq
\label{Lpn1}
{\cal L}_{\pi N}^{(1)}= \overline{\Psi}(i\sla{D}-m+
\frac{1}{2}g_A \sla{u} \gamma_5)\Psi \, .
\eeq
Here the pion and nucleon fields are collected as
\beq u^2=U=e^{i \tdpi/\F}, \quad \quad 
\Psi= \left( \begin{array}{c} p\\n \end{array} \right)  \, . \eeq
The covariant derivative $D_\mu$, when acting on things transforming 
as nucleon fields, is defined as
\beq D_\mu \Psi = (\dv _\mu + \Gamma_\mu) \Psi   \, , \eeq
with
\beq \Gamma_\mu = \frac{1}{2} [u^\dag(\dlmu-i r_\mu) u + 
u(\dlmu-i \ell_\mu)u^\dag] \eeq
and with $\ell_\mu$ and $r_\mu$ constructed from the external vector 
and axial vector currents as $\ell_\mu=v_\mu-a_\mu$ and $r_\mu=v_\mu+a_\mu$.
We also have
\beq u_\mu = i [u^\dag(\dlmu-i r_\mu) u - 
u(\dlmu-i \ell_\mu)u^\dag] \, . \eeq
The parameters appearing in this lowest order Lagrangian, $m,F_0,g_A$
are respectively the 'bare' or unrenormalized values of the nucleon
mass, the pion decay coupling and $G_A(0)$ and the fields are the
'bare' fields.

The higher order Lagrangians are given by
\beqa
\label{Lpn234}
{\cal L}_{\pi N}^{(2)} =
\sum_{i=1}^7 c_i\overline{\Psi}{\cal O}_i\Psi\, , 
\ \ \
{\cal L}_{\pi N}^{(3)}=
\sum_{i=1}^{23} d_i\overline{\Psi}{\cal O}_i\Psi\, ,
\ \ \
{\cal L}_{\pi N}^{(4)}=
\sum_{i=1}^{118} e_i\overline{\Psi}{\cal O}_i\Psi \, ,
\eeqa
where the $c_i, d_i, e_i$ are the LEC's and where the basis functions
${\cal O}_i$ are given in Ref.~\cite{Fettes00} Tables III, IV and
V. Note however that we always normalize the LEC's to the physical
mass $m_N$ rather than to $m$.

We will also need the Lagrangian in the purely meson sector, as the 
results depend on that choice as well. We will take the standard one
\beq 
{\cal L}_{\pi } = {\cal L}_{\pi }^{(2)}+{\cal L}_{\pi }^{(4)}+\cdots \, ,
\eeq
where the lowest order Lagrangian is given by
\beq
\label{Lp2}
{\cal L}_{\pi }^{(2)}=\frac{F_0^2}{4} <D_{\mu}U (D^{\mu}U)^{\dagger}>+
\frac{F_0^2}{4}<\chi U^\dagger+ U\chi^\dagger> .
\eeq
Here the covariant derivative acting on quantities transforming as 
$U$ is given by
\beq
D_\mu U = \dlmu U -i r_\mu U + i U \ell_\mu \, . 
\eeq
We also have 
\beq
\chi=2B_0 (s + ip)\, ,
\eeq
where $s$ and $p$ are the external scalar and pseudoscalar current and
the chiral symmetry breaking is introduced as usual by taking $p=0$
and $s=\hat{m}$ with $\hat{m}=(m_u+m_d)/2$ the average of up and down
quark masses. The parameter $B_0$ is given in terms of the lowest order 
pion mass $m_{0\pi}$ (cf. Eq.~(\ref{physpimass})) by 
$2 B_0\hat{m}=m_{0\pi}^2$.  For the fourth order Lagrangian
we take the Gasser-Sainio-\v{S}varc \cite{Gasser} form of the
Gasser-Leutwyler Lagrangian \cite{GL} given explicitly for example by
Eq.~(D.13) of Ref.~\cite{Srev}.

\subsection{Counting rules and 'non counting' terms from loop diagrams} 

In the usual HBChPT approach the Lagrangians ${\cal L}_{\pi N}^{(1)},
{\cal L}_{\pi N}^{(2)}, {\cal L}_{\pi N}^{(3)}, {\cal L}_{\pi
N}^{(4)}$, or more precisely the HBChPT expansions of these
Lagrangians, contribute tree level diagrams respectively of order
$p,p^2,p^3,p^4$ where we mean by $p$ the generic small expansion
parameter, e.g., $m_\pi/m_N$. As a consequence of using the dimensional
regularization procedure for regularizing the integrals, \cite{Djuk05}
the one loop graphs containing only vertices from ${\cal L}_{\pi
N}^{(1)}$ contribute at ${\cal O}(p^3)$ and those containing one
vertex from ${\cal L}_{\pi N}^{(2)}$ contribute at ${\cal
O}(p^4)$. Two or more loop graphs contribute only at ${\cal O}(p^5)$
or higher.

In the relativistic approach however the counting breaks down, with
multi-loop graphs contributing to lower orders, ${\cal O}(p^2)$,
${\cal O}(p^3)$, etc.~\cite{Gasser}, than that obtained in
HBChPT. Thus one needs to develop some different scheme for ordering
the various contributions and determining which to keep.

From a purely practical point of view it is rarely possible, or
necessary, to consider more than one loop. Furthermore one of the most
important general results of Becher and Leutwyler \cite{BL} or more
particularly Fuchs, {\it et al.} \cite{fuchs03} is that it is always
possible to absorb into the LEC's those terms from multi-loop diagrams
which do not obey the HBChPT counting rules. Thus we will here
consider only one loop diagrams and assume that all contributions from
multi-loop diagrams are either ${\cal O}(p^5)$ or, in accord with the
general result, have been absorbed in the LEC's in the original
Lagrangians as defined in Eqs.~(\ref{Lpn1}) and (\ref{Lpn234}).

We note also, as will be seen from the explicit expressions below,
that associated with each loop is the factor $1/(4\pi \F)^2$. By
neglecting multi-loop diagrams we are neglecting terms containing
higher powers of this factor, e.g. $1/(4\pi \F)^4$. Such higher powers
will also arise from the expansion of unrenormalized quantities in one
loop diagrams about renormalized values. We will always drop such
higher powers of $1/(4\pi \F)^2$, arguing that such approximation is
consistent with our neglect of multi-loop diagrams.

Thus to summarize, we will work consistently to one loop, and to
${\cal O}(p^4)$. This means that we will keep all terms of ${\cal
O}(p^4)$ or lower except for those originating in multi-loop diagrams,
which we assume to have been absorbed in the LEC's. We also drop terms
involving $1/(4\pi \F)$ to the fourth or higher power, an
approximation which is consistent with the neglect of multi-loop
contributions, which have these same factors. Note also that those one
loop diagrams which have higher order Lagrangians at the vertices and
thus which are of ${\cal O}(p^5)$ or higher in the HBChPT sense will
also be dropped, again assuming that any lower order terms from these
diagrams which do not obey this counting have been absorbed in the
LEC's.

This choice of diagrams and terms to keep is to some extent dictated
by the practicalities of doing such calculations. Two loop diagrams
and those with many higher order vertices are difficult to handle, and
normally would be considered only if there were some special
circumstance which suggests that they would be large. Perhaps the most
important result arising from the work of Refs.~\cite{BL,fuchs03} is
that this is a consistent procedure, i.e. that lower order
contributions which originate in these higher order diagrams which we
must neglect can in fact be absorbed in the LEC's in a way that
preserves the usual HBChPT counting procedures.

In the relativistic approach there will also be terms, a finite number
of them, coming from the one loop diagrams we keep, which do not obey
HBChPT counting, e.g. those of ${\cal O}(p^2)$. We will flag these
terms, but for now keep all of them explicitly, until we discuss the
various renormalization schemes which have been proposed, as these
schemes differ primarily in how they treat these 'non counting' terms.

\section{Evaluation of the NN 3-point vertex : vector and axial 
vector currents}
\subsection{Preliminaries}
We now proceed to evaluate the amplitude of Eq.~(\ref{OMCamp}). The
amplitude originates in the usual current-current interaction which
couples the lepton current to the weak nucleon current. The lepton
current is given by the relativistic tree-level current,
$l_\alpha=\bar{u}_\nu\gamma_\alpha(1-\gamma_5)u_\mu$. The $(V-A)$ weak
nucleon current is calculated from the effective Lagrangian using the
approach described above. 

There are three contributions to the weak-nucleon-nucleon vertex. The
first two of these involve coupling of the nucleons to an external
vector field and to an external axial field. The diagrams which
contribute to these two contributions are given in
Fig.~\ref{fig:avdiags}.  The third contribution, to be discussed in
the next section, involves coupling to the pion, which by virtue of
its coupling to the leptonic current contributes to the overall axial
weak nucleon current.

We take for the external vector current $v_\mu \rightarrow
v_\mu^{(s)}+ \tdvv_\mu$, i.e., we divide the current into isoscalar
and isovector part. Only the isovector part contributes to the weak
current, but we will keep both for completeness, and to allow
evaluation of some of the LEC's via a connection to the isoscalar
electromagnetic form factors of the nucleon. Similarly we take $a_\mu
\rightarrow a_\mu^{(s)}+ \tdva_\mu$, but in this case will drop the
isoscalar axial current $a_\mu^{(s)}$.

\subsection{Tree level diagrams}

The tree level contributions to the amplitude correspond to diagrams
1-4 in Fig.~\ref{fig:avdiags} and are given by
\beqa
M_{1V}&=& i\sqrt{Z_N} \Psibn (v_\mu^{(s)}+\tdvv_\mu) \gamma^\mu 
\Psip \sqrt{Z_N} \, ,  \label{eq:M1V} \\
M_{1A}&=& i g_A \sqrt{Z_N} \Psibn \tdva_\mu \gamma^\mu \gamma_5 
\Psip\sqrt{Z_N}  \, ,  \label{eq:M1A} \\
M_{2V}&=& i\sqrt{Z_N} \Psibn \frac{i \sigma^{\mu\nu} q_\nu}{2 m_N}[(c_6+
2 c_7) v_\mu^{(s)}+c_6 \tdvv_\mu] \Psip \sqrt{Z_N}  \, ,\label{eq:M2V} \\
M_{2A}&=& 0\, , \\
M_{3V}&=& i \Psibn (q^\mu q^\nu - q^2 g^{\mu\nu})(p_i+p_f)_\nu(
\frac{2 d_7}{m_N} v_\mu^{(s)}+\frac{d_6}{m_N} \tdvv_\mu) \Psip \, , \\
M_{3A}&=& i \Psibn \gamma^\mu \gamma_5 [4 m_{0\pi}^2 d_{16} \tdva_\mu + 
d_{22} (q^2 g_{\mu\nu}-q_\mu q_\nu) \tdva^\nu]\Psip \, , \\
M_{4V}&=& i \Psibn i \sigma^{\mu\nu}q_\nu [4(q^2 e_{54}-
4 m_{0\pi}^2 e_{105}) v_\mu^{(s)}+2(q^2 e_{74}-
4 m_{0\pi}^2 e_{106})\tdvv_\mu] \Psip\, , \\
M_{4A}&=& 0 \, .
\eeqa
Here the subscript $V$ or $A$ refers to coupling to vector or axial
current respectively and the number refers to the particular diagram
in Fig.~\ref{fig:avdiags}. For these and subsequent amplitude
expressions $\Psibn$ and $\Psip$ are to be interpreted as wave 
functions, rather than the fields of the original Lagrangian. That is,
they are still two component objects in isospin space, but made up of
spinors $\overline{u}(p_f)$ and $u(p_i)$ rather than fields. In the
HBChPT counting system these tree level diagrams contribute to ${\cal
O}(p)$, ${\cal O}(p^2)$, ${\cal O}(p^3)$,${\cal O}(p^4)$
respectively. Note that the nucleon wave function renormalization
factor $Z_N$  appears only in the $M_1$ and $M_2$ amplitudes which is a
consequence of the fact, as shown in Eq.~(\ref{eq:ZN}) of 
Appendix \ref{ap:massren}, that
the leading corrections to $Z_N$ are two orders higher, $Z_N=1+{\cal
O}(p^2)$.

\subsection{Leading one loop diagrams}
The next set of diagrams consists of those one loop diagrams with all
vertices coming from ${\cal L}_{\pi N}^{(1)}$. In the HBChPT sense
these all contribute first at ${\cal O}(p^3)$ , but in the relativistic
approach they will also contribute  some 'non counting' terms of 
${\cal O}(p^2)$ as well as relativistic corrections of ${\cal O}(p^4)$ 
and higher.

We express these amplitudes in terms of a general loop integral
$I_{\pi\pi....NN....}$ which is defined in detail in Appendix
\ref{ap:loopinteg}. For present purposes it is sufficient to note that
$I_{\pi\pi....NN....}[k_i,....p_j,...,A]$ refers to a loop integral
which contains a pion propagator for each subscript $\pi$ with momenta
of the form $k_i+\ell$ and a nucleon propagator denominator for each
subscript $N$ with momenta of the form $p_j+\ell$.  The loop
integration variable $\ell$ \cite{footnote2} is chosen so that the first
pion momentum $k_1$ is zero and that argument is normally not put in
explicitly. $A$ is whatever is in the numerator.

In terms of these loop integrals the lowest order one loop amplitudes
are given by
\beqa
M_{5V}&=& \frac{i g_A^2}{4 \F^2}\Psibn (3 v_\mu^{(s)}-\tdvv_\mu)
I_{\pi N N}[p_i,p_f,
\sla{\ell}\gamma_5(\sla{p}_f+\sla{\ell}+m)\gamma^\mu(
\sla{p}_i+\sla{\ell}+m)\sla{\ell}\gamma_5] \Psip \, , \\
M_{5A}&=&  -\frac{i g_A^3}{4 \F^2} \Psibn \tdva_\mu I_{\pi N N}[p_i,p_f,
\sla{\ell}\gamma_5(\sla{p}_f+\sla{\ell}+m)\gamma^\mu\gamma_5(
\sla{p}_i+\sla{\ell}+m)\sla{\ell}\gamma_5] \Psip \, , \\
M_{8V}&=&\frac{i g_A^2}{\F^2} \Psibn \tdvv_\mu\{I_{\pi N}[p_f,
\sla{\ell}\gamma_5(\sla{p}_f+\sla{\ell}+m)\gamma^\mu\gamma_5] \nn \\
& & \eqindent +I_{\pi N}[p_i,
\gamma^\mu\gamma_5(\sla{p}_i+\sla{\ell}+m)\sla{\ell}\gamma_5]\}\Psip\, , \\
M_{8A}&=&\frac{i g_A}{\F^2} \Psibn \tdva_\mu\{I_{\pi N}[p_f,
\sla{\ell}\gamma_5(\sla{p}_f+\sla{\ell}+m)\gamma^\mu]+I_{\pi N}[p_i,
\gamma^\mu(\sla{p}_i+\sla{\ell}+m)\sla{\ell}\gamma_5]\}\Psip \, , \\
M_{11V}&=& - \frac{i}{\F^2}\Psibn \tdvv_\mu \gamma^\mu \Psip I_\pi[1]\, , \\
M_{11A}&=& - \frac{i g_A}{\F^2}\Psibn \tdva_\mu \gamma^\mu \gamma_5 
\Psip I_\pi[1]\, , \\
M_{13V}&=& -\frac{i g_A^2}{\F^2}\Psibn \tdvv_\mu I_{\pi \pi N}[-q,p_i,
(2 \ell-q)^\mu(\sla{\ell}-\sla{q})\gamma_5(\sla{p}_i+\sla{\ell}+m)
\sla{\ell}\gamma_5]\Psip\, , \\
M_{13A}&=& 0\, , \\
M_{15V}&=& \frac{i}{2\F^2}\Psibn \tdvv_\mu  
I_{\pi\pi}[-q,(2 \ell-q)^\mu(2 \sla{\ell}-\sla{q})]\Psip \, , \\
M_{15A}&=& 0 \, .
\eeqa

\subsection{Further one loop diagrams}
The next class of one loop diagrams consists of those with one vertex
from ${\cal L}_{\pi N}^{(2)}$ with all others from ${\cal L}_{\pi
N}^{(1)}$. In the HBChPT sense these would be of ${\cal O}(p^4)$, but
again in the relativistic formulation they will have 'non counting' terms
of lower order.

These amplitudes are given by
\beqa
M_{6V}&=&\frac{i g_A^2}{4 \F^2} \Psibn [3(c_6+2 c_7)v_\mu^{(s)}-
c_6 \tdvv_\mu] \nn \\
& & \eqindent  \times I_{\pi N N}[p_i,p_f,\sla{\ell}\gamma_5(\sla{p}_f+
\sla{\ell}+m)\frac{i \sigma^{\mu\nu}q_\nu}{2 m_N}(\sla{p}_i+
\sla{\ell}+m)\sla{\ell}\gamma_5]\Psip\, , \\
M_{6A}&=& 0\, , \\
M_{9V}&=& 0\, , \\
M_{9A}&=& - \frac{i g_A}{4 \F^2}\Psibn \tdva_\mu \left\{
I_{\pi N}[p_f,\sla{\ell}\gamma_5(\sla{p}_f+\sla{\ell}+m)
[\frac{4 c_2}{m_N^2}(p_i^\mu p_i \cdot \ell +
(p_f+\ell)^\mu (p_f+\ell)\cdot \ell) \right.  \nn \\ 
&+& \left. 8 c_3 \ell^\mu -
2 i \sigma^{\mu\nu}(4 c_4 \ell_\nu+\frac{c_6}{m_N} q_\nu)]] \right. 
\nn \\ &+&
\left. I_{\pi N}[p_i,[\frac{4 c_2}{m_N^2}(p_f^\mu p_f \cdot \ell +
(p_i+\ell)^\mu (p_i+\ell)\cdot \ell)+8 c_3 \ell^\mu \right. 
\nn \\&+& \left.
2 i \sigma^{\mu\nu}(4 c_4 \ell_\nu-\frac{c_6}{m_N} q_\nu)]
(\sla{p}_i+\sla{\ell}+m)\sla{\ell}\gamma_5]\right\} \Psip \, , \\
M_{12V}&=& \frac{3 i c_2}{\F^2 m_N^2} \Psibn (v_\mu^{(s)}+\tdvv_\mu)
(p_i+p_f)_\nu I_\pi[\ell^\mu \ell^\nu] \Psip+
\frac{c_6}{2 \F^2 m_N} \Psibn \sigma^{\mu\nu} q_\nu \tdvv_\mu 
\Psip I_\pi[1]\, , \\
M_{12A}&=& 0\, , \\
M_{16V}&=& -\frac{2 c_4}{\F^2} \Psibn \sigma^{\mu\nu} q_\nu \tdvv_\alpha 
I_{\pi\pi}[-q,(2 \ell-q)^\alpha \ell_\mu]\Psip\, , \\
M_{16A}&=& 0 \, .
\eeqa

\subsection{Mass insertion terms}
The final set of amplitudes arises from the mass insertions on
internal nucleon lines. These insertions come from the NN two point
term in ${\cal L}_{\pi N}^{(2)}$, namely $\Psibn 4 c_1
m_{0\pi}^2 \Psip$. The relevant amplitudes, those for diagrams 7,10,14 
of Fig.~\ref{fig:avdiags},
can be obtained from the underlying diagrams 5,8,13 respectively by
the substitution for each nucleon propagator in turn
\beq 
\label{propreplace}
\frac{i}{\sla{p}- m +i \epsilon} \rightarrow 
\frac{i}{\sla{p}- m +i \epsilon}(4 i c_1 m_{0\pi}^2)
\frac{i}{\sla{p}- m +i \epsilon}.
\eeq

An alternative procedure is to observe that $m$ appears in the loop
integrals only in the propagators, since we normalized all the
constants in the Lagrangian to $m_N$ not $m$. Thus we can use the fact
that
\beq 
i(4 i c_1 m_{0\pi}^2)\frac{\partial}{\partial m} ( \frac{i}{\sla{p}- 
m +i \epsilon} ) = \frac{i}{\sla{p}- m +i \epsilon}(4 i c_1 m_{0\pi}^2)
\frac{i}{\sla{p}- m +i \epsilon}.
\eeq
This allows us to obtain the amplitudes with mass insertions by taking
derivatives of the corresponding amplitudes without insertions. As
detailed in Appendix \ref{ap:loopinteg} this approach, while exact,
may not be as useful as one might expect because the derivative in
effect reduces the power of small expansion parameter. Thus in some
cases one has to expand the initial integral to higher order than
needed for it alone so as to get the mass insertion diagrams to the
appropriate order. We actually calculated these mass insertion
diagrams explicitly and used this derivative procedure to check the
results.

Another approach is to observe that (see Appendix \ref{ap:massren})
the 'bare' nucleon mass which appears in the original Lagrangian is
related to the physical mass by the relation
\beq
m_N=m - 4 c_1 m_{0\pi}^2+{\cal O}(p^4).
\eeq
Thus a propagator with mass $m \rightarrow m_N$ can be expanded as
\beq
\label{propexp}
\frac{i}{\sla{p}- m_N +i \epsilon} \rightarrow 
\frac{i}{\sla{p}- m +i \epsilon}+
\frac{i}{\sla{p}- m +i \epsilon}(4 i c_1 m_{0\pi}^2)
\frac{i}{\sla{p}- m +i \epsilon}+... \, .
\eeq

Thus if we replace $m \rightarrow m_N$ in the propagators in the one
loop diagrams we effectively include the mass insertion diagrams
7,10,14 of Fig.~\ref{fig:avdiags}. At the same time we reduce the
number of separate diagrams to be calculated and reduce by one the
maximum number of propagators involved in the loop integrals which
must be calculated. Both of these offer significant calculational
advantages. However this expansion only works to first order, so one
must keep only terms in the expansion of the propagators which are
linear in $c_1$. This requires extreme care since $c_1$ appears also
in other places in the amplitudes so that in general there are
legitimate terms involving $c_1^2$ which must be kept.

Note that simply replacing $m \rightarrow m_N$ in all the propagators
is not exactly equivalent to the direct calculation of the mass
insertion diagrams. The expansion of Eq.~(\ref{propexp}) only works to
lowest order, so with such a substitution there will be some spurious
higher order terms implicitly included. Also implicitly included will
be diagrams like those of Fig.~\ref{fig:highord} which involve two
mass insertions or which involve one mass insertion plus a vertex from
${\cal L}_{\pi N}^{(2)}$. These would not have been included in a
direct calculation of the mass insertion diagrams as they would have
been nominally of too high order. To the order we are considering most
of these extra terms can be neglected. In fact in HBChPT all would be
of higher order.  However in the relativistic approach there will be a
few terms arising from the 'non counting' terms from diagrams such as
those of Fig.~\ref{fig:highord} which will appear in the amplitude in
this approximation and not in the explicit calculation.

To repeat, our approach was to calculate the mass insertion diagrams
explicitly and thus our results may differ from calculations which use
one of the above approximations.

\subsection{Summary}
By summing all of the amplitudes given above  we
obtain the complete weak nucleon-nucleon amplitude arising from
interaction with external vector and axial vector fields.

\section{Evaluation of the NN 3-point vertex : pion terms}
The third contribution which must be evaluated comes from the
pion-nucleon-nucleon vertex and will contribute to the axial
current. The diagrams which are needed are given in
Fig.~\ref{fig:pnndiags} and the amplitudes associated with those
diagrams are given by
\beqa
M_{1\pi}&=&-\frac{g_A}{2 \F} \sqrt{Z_N} \Psibn \tdpi \sla{q} \gamma_5 
\Psip \sqrt{Z_N} \sqrt{Z_\pi} \, , \label{eq:M1pi} \\
M_{2\pi}&=& 0 \, , \\
M_{3\pi}&=& \frac{m_{0\pi}^2}{\F} (d_{18}-2 d_{16}) 
\Psibn \tdpi \sla{q} \gamma_5 \Psip\, , \\
M_{4\pi}&=& 0 \, , \\
M_{5\pi}&=& \frac{g_A^3}{8 \F^3}\Psibn \tdpi I_{\pi NN}[p_i,p_f,
\sla{\ell} \gamma_5 (\sla{p_f}+\sla{\ell}+m)
\sla{q}\gamma_5(\sla{p_i}+\sla{\ell}+m)\sla{\ell}\gamma_5]\Psip\, , \\
M_{6\pi}&=& 0 \, , \\
M_{8\pi}&=&-\frac{g_A}{4 \F^3}\Psibn \tdpi \left\{ I_{\pi N}[p_f,
\sla{\ell}\gamma_5(\sla{p_f}+\sla{\ell}+m)(\sla{q}-\sla{\ell})]\right. 
\nn \\ & & + \left.
I_{\pi N}[p_i,(\sla{q}+\sla{\ell})(\sla{p_i}+\sla{\ell}+m)
\sla{\ell}\gamma_5]\right\}\Psip\, , \\
M_{9\pi}&=& -\frac{g_A}{2 \F^3}\Psibn \tdpi \{
I_{\pi N}[p_f,\sla{\ell}\gamma_5 (\sla{p_f}+\sla{\ell}+m)[-4 c_1 
m_{0\pi}^2 \nn \\ 
& &-\frac{c_2}{m_N^2}((p_f+\ell)\cdot \ell (p_f+\ell)\cdot q +
p_i \cdot \ell p_i \cdot q)-2 c_3 \ell \cdot q -
2 i c_4 \sigma ^{\mu\nu} \ell_\mu q_\nu]] \nn \\ & & -
I_{\pi N}[p_i,[-4 c_1 m_{0\pi}^2 +\frac{c_2}{m_N^2}((p_i+\ell)\cdot \ell 
(p_i+\ell)\cdot q +p_f \cdot \ell p_f \cdot q)+2 c_3 \ell \cdot q 
\nn \\ & & -
2 i c_4 \sigma ^{\mu\nu} \ell_\mu q_\nu](\sla{p_i}+\sla{\ell}+m)
\sla{\ell}\gamma_5]\} \Psip
\, , \\
M_{11\pi}&=& \frac{g_A}{6 \F^3}\Psibn\tdpi\sla{q}\gamma_5 
\Psip I_\pi[1]\, , \\
M_{12\pi}&=& 0 \, . \label{eq:M12pi}
\eeqa

Again the mass insertion diagrams 7 and 10 are obtained by making the
replacement of Eq.~(\ref{propreplace}) in diagrams 5 and 8.

To get the contribution to the weak nucleon-nucleon axial current from
this $\pi NN$ amplitude we make the replacement (for $q^2 \neq
m_\pi^2$) $\tdpi \rightarrow 2 i \Fpi q^\mu \tdva_\mu/(q^2-m_\pi^2)$.
This arises from the addition of a pion propagator and pion decay
vertex to the $\pi NN$ vertex. Note that the parameters $\Fpi$ and
$m_\pi$ are the physical ones. Since we have associated a
$\sqrt{Z_\pi}$ with the amplitude $M_{1\pi}$ so that it is
renormalized, we need to use here the renormalized propagator and
renormalized pion decay vertex to make the overall amplitude
renormalized.

\section{Evaluation, regularization and renormalization}
The first step in evaluating these amplitudes is to reduce the
numerators of the loop integrals. This is done using the standard
algebra of Dirac matrices and the usual tensor decomposition of
integrals with $\ell^\mu, \ell^\mu \ell^\nu, ...$ in the
numerator. The end result is that the full amplitude can be expressed
in terms of the following loop integrals with unit numerator:
$I_\pi[1],I_{\pi N}[p,1],I_{NN}[p_i,p_f,1],I_{\pi \pi}[q,1],I_{\pi
NN}[p_i,p_f,1],I_{\pi\pi N}[-q,p,1]$, where $p$ can be $p_i$ or
$p_f$. For the diagrams with the mass insertion put in explicitly we
need the additional integrals $I_{\pi
NN}[p_i,p_i,1],I_{NN}[p_i,p_i,1],I_{NNN}[p_i,p_i,p_f,1],I_{\pi\pi
NN}[-q,p_i,p_i,1]$ plus the corresponding ones with the roles of $p_i$
and $p_f$ interchanged.

In HBChPT these integrals are evaluated using dimensional
regularization to extract the divergences, which are then absorbed in
the renormalization of the LEC's. In the relativistic approach this
procedure works in the same way for integrals involving only pion
propagators. Thus for example we have in standard fashion for
dimension $d \simeq 4$
\beq
I_\pi[1]=\frac{m_{0\pi}^2}{(4 \pi)^2}[R+{\rm ln}(\frac{m_{0\pi}^2}{\mu^2})] 
\eeq
where
\beq \label{Rdef} 
R=-\frac{1}{\epsilon}+\gamma-1-{\rm ln}(4\pi), \quad \epsilon = 
\frac{4-d}{2}
\eeq
with $\gamma=-\Gamma'(1)=0.577....$.

In the relativistic approach the same procedure applied to integrals
containing nucleon propagators leads to the $\widetilde{MS}$ scheme in
which the $R$'s are all absorbed in the LEC's. The amplitudes however
still contain finite terms which do not obey the usual counting rules
of HBChPT. Thus for example one loop diagrams which are nominally
${\cal O}(p^3)$ may contain contributions at
${\cal O}(p^2)$ and likewise those nominally ${\cal O}(p^4)$ 
may contain also terms of ${\cal O}(p^2)$ and ${\cal O}(p^3)$.  
There have been two somewhat
different, but similar, methods proposed to resolve this problem. In
the 'Infrared Renormalization' (IR) scheme proposed by Becher and
Leutwyler \cite{BL} the loop integrals are divided into two parts. An
'infrared singular' part contains non integer powers of the small
expansion parameter and a 'regular' part contains only integer
powers. They then renormalize the integrals by dropping the regular
part and absorbing the infinities of the singular part, i.e. the terms
proportional to $R$, in a renormalization of the LEC's. Note that
'drop' means 'absorb in the LEC's' or equivalently 'cancel via
counterterms in the Lagrangian'. Thus in this approach the infinities
which appear in both singular and regular parts are in effect combined
and absorbed in the LEC's in the same way as would be done in the
$\widetilde{MS}$ scheme. The difference arises in that in the IR
scheme additional regular polynomial terms, including all those which
do not satisfy usual HBChPT counting, are also absorbed in the LEC's.

The other approach, the Extended On Mass Shell (EOMS) scheme of the
Mainz group \cite{fuchs03}, first uses the usual dimensional
regularization to extract the terms proportional to $R$, which are then
absorbed in renormalizations of the LEC's in exactly the same fashion
as in the $\widetilde{MS}$ scheme. In the second step the amplitude
for each individual diagram is examined and those terms, all
polynomials in the expansion parameter, which do not obey the counting
rules, as used in HBChPT, are determined. This finite set of terms,
the 'non-counting' terms, are then dropped, i.e. absorbed into the
LEC's. This approach thus eliminates a somewhat smaller set of terms
than does the IR approach.

In practical applications the EOMS approach involves a number of
subtleties. These subtleties, basically amounting to choices of 
conventions, affect the specific terms absorbed in the LEC's and
thus make little difference as long as one is considering just one
process. They simply change slightly the numerical values of the
LEC's. However if one wants a consistent scheme to be applied to a
variety of processes, as is our intent here, it is necessary to
discuss the various choices and to define exactly what conventions 
we take, as one
must use the same conventions in subsequent calculations or when using
values of the LEC's extracted by others.

First, when extracting the non-counting terms from each diagram it
makes a difference whether one first expresses the amplitudes as a
function of the original mass $m$ appearing in the Lagrangian, the 
mass in the
chiral limit $\stackrel{\circ}{m}$, or the physical mass
$m_N$. Differences are ${\cal O}(m_\pi^2)$ and would thus be higher
order corrections in the HBChPT scheme. In the relativistic approach 
however such corrections to, say, ${\cal O}(p^2)$ non-counting terms 
can enter as ${\cal O}(p^4)$ terms 
which are kept. Thus absorbing a non-counting term expressed as a
function of $m$ leads to a slightly different renormalization than
would be obtained by absorbing the equivalent term expressed as a
function of $m_N$. Here we always express the amplitudes in terms of
the physical mass $m_N$ before isolating the non-counting terms.

Similarly the log terms in the amplitudes can be expressed as ${\rm
ln}(m^2/\mu^2)$ which makes $\mu=m$ the logical choice, since then 
these terms vanish,  or as ${\rm ln}(m_N^2/\mu^2)$ 
which makes $\mu=m_N$ the logical choice. We have
kept the scale parameter $\mu$ explicit until the end, but have used
$m_N$ in the amplitudes and eventually for $\mu$.

An essentially similar effect arises in the approach used for
including the diagrams with mass insertions on internal nucleon
lines. We evaluated each diagram explicitly so that for each of the
${\cal O}(p^3)$ diagrams with an internal nucleon line, i.~e.,
Fig.~\ref{fig:avdiags} diagrams 5,8,13 and Fig.~\ref{fig:pnndiags}
diagrams 5,8, there is an associated diagram of ${\cal O}(p^4)$,
Fig.~\ref{fig:avdiags} diagrams 7,10,14 and Fig.~\ref{fig:pnndiags}
diagrams 7,10 respectively. We then looked at each diagram in the two
sets to determine the non-counting terms for each. An alternative
approach used in \cite{fgs-jpg04} replaces $m \rightarrow m_N$ in the
${\cal O}(p^3)$ diagrams and then later expands to first order in the
difference to get the mass insertion contributions. In this approach a
term of ${\cal O}(p^2)$ before expansion would be dropped, as the
original diagrams are nominally of ${\cal O}(p^3)$. However had the
expansion been done first, the expansion of such terms would give
pieces of ${\cal O}(p^4)$ which one would keep, and which in fact are
some of those arising in the diagrams with explicit mass insertions.

Finally observe that, although the EOMS scheme is based on extracting
from each individual diagram those terms which do not obey the nominal
counting rules and absorbing those terms in LEC's, that procedure does
not ensure that each individual diagram obeys the counting rules.  The
exact same statement can be made for the IR scheme and a very similar
statement can be made for the $\widetilde{MS}$ scheme, where the
renormalization of the LEC's to remove the infinities ensures that the
amplitude is finite, but not that each individual diagram is
finite. In all these cases the situation occurs because in general
there are not always LEC's available to absorb terms from individual
diagrams. The simplest example of this can be seen in the calculation
of the $q^2=0$ limit of $G_V$. In the EOMS scheme explicit calculation
shows that the diagrams 7,8,10,13, and 14 of Fig.~\ref{fig:avdiags}
all contribute non-counting terms. Many of these are removed by the
renormalization provided by $Z_N$ (appearing in the tree level
diagrams 1 and 2) as given in Eq.~(\ref{eq:ZN}) of
Appendix~\ref{ap:massren}. However two contributions remain, those
from diagrams 8 and 13. There are no LEC's available here and so no
way to absorb these as individual terms. Instead what happens is that
these two contributions cancel each other so that the sum of the
amplitudes from all individual diagrams contains no non-counting
terms.

Similarly in the $\widetilde{MS}$ scheme diagrams 5,7,8,10,11,13,14,
and 15 of Fig.~\ref{fig:avdiags} all contain infinite terms
proportional to $R$. Again $Z_N$ renormalizes some of these away, but
there are a number of terms left and no available LEC's to absorb
them. Instead they cancel among themselves.

Naively it is perhaps obvious that this happens as individual diagrams
are not physically measurable quantities and thus do not necessarily
satisfy physically relevant constraints such as counting or
finiteness. More precisely, the argument that the finite number of
non-counting terms, or the infinities, can be absorbed in the LEC's
(or in counterterms) relies on the fact that the Lagrangian contains
all possible counterterms allowed by the symmetries. In this case the
relevant symmetry is current conservation which, as is well known,
ensures that the weak vector coupling (or the isovector
electromagnetic coupling $F_1$) is not renormalized by the strong
interactions. This symmetry is obeyed by the full amplitude, but not
by individual diagrams. Thus it perhaps should not be a surprise that
terms are generated for individual diagrams which cannot individually
be absorbed in counterterms. The fact that the sum cancels as required
however is a clear check on the correctness of the calculation.

Thus to summarize we consider three approaches, the $\widetilde{MS}$,
IR and EOMS schemes. In all three one first regularizes the integrals
using dimensional regularization and renormalizes the LEC's to remove
all infinities in the usual way. For IR and EOMS additional sets of
finite terms are extracted and absorbed in the LEC's so as to preserve
counting. The difference between the two is simply in the explicit
terms extracted.

Note that these three approaches will not lead to any different
predictions for measurable quantities. The formulas for such
quantities in terms of the LEC's will be different, but that will be
compensated by different formulas for the LEC's in terms of
unrenormalized quantities and different numerical values for the
LEC's.

\section{Results}

We thus proceed as outlined above, i.e. we extract $R$ from each of the
remaining loop integrals using the standard dimensional regularization
as in Appendix {\ref{ap:loopinteg}. The integrals are first separated
into two parts according to the IR prescription. The parts involving $R$
are recombined and the renormalization of the LEC's to absorb $R$
proceeds just as in the usual $\widetilde{MS}$ scheme. Thus the
original LEC's in the Lagrangian, $x$, where $x$ stands for any of the
LEC's, are eliminated in favor of their $\widetilde{MS}$ renormalized
values $x^r$. The finite parts are then expanded in powers of the
small parameter and terms through ${\cal O}(p^4)$ are kept. We flag
those terms originating in the infrared regular part of the integral
with a parameter $\BL$, as detailed in Appendix {\ref{ap:loopinteg}. 
We assume that $q^2/m_\pi^2 \leq 1$, but not
necessarily very small, but for simplicity keep only terms linear in
$q^2$ in the final results. We also replace the original parameters of
the Lagrangian, $m, m_{0\pi}, \F$ with their physical values as
determined in Appendix \ref{ap:massren}. Then the contribution of each
diagram is examined and those terms which do not obey counting and
which would be dropped in the EOMS scheme are flagged with a parameter
$\beta^{EOMS}$.

The full amplitude is then evaluated and put in the form of
Eq.~(\ref{OMCamp}) which gives the vector and axial vector weak
nucleon currents, $V^\mu, A^\mu$, appropriate for muon capture and
allows us to extract explicit expressions for the various form factors
in the equation.  Two further renormalizations are then performed. The
first expresses $x^r$ in terms of the EOMS renormalized quantities
$x^{EOMS}$ and is determined by requiring that all terms flagged by
$\beta^{EOMS}$ must be absorbed. The second expresses $x^{EOMS}$ in
terms of the IR renormalized LEC's $x^{IR}$ and removes all the
remaining terms proportional to $\BL$. The expressions for the
renormalizations seem to be unique, except for the few cases where
only a combination of LEC's appears, as long as each renormalization
involves only terms with the same power of $m_\pi$.

The weak form factors expressed in terms of the IR renormalized LEC's
are given by the following. In these relations we have always used
physically measurable masses, $m_N, m_\pi$ and coupling $\Fpi$, have
taken the scale factor $\mu \rightarrow m_N$, and have kept only terms
linear in $q^2$, but otherwise have kept all terms consistent with
an expansion of the amplitude to fourth order in the expansion in the
small parameter.

\beqa 
G_V(q^2)&=&1+q^2\left(-2 d_6^{IR}- \frac{1}{96 \FPsq}(7
g_A^2+1+(5 g_A^2+1) \lnmpi)\right. \nn \label{res:GV} \\ & & +
\left. \frac{35 g_A^2 m_\pi}{192 \Fpi^2 \pi m_N}\right) \, , \\
G_M(q^2)&=&c_6^{IR} -
\frac{g_A^2 m_N m_\pi}{4 \Fpi^2 \pi}-
16 e_{106}^{IR} m_N m_\pi^2 -
\frac{g_A^2 m_\pi^2}{32 \FPsq}(4 c_6 +8) \nn \\ & & -
\frac{m_\pi^2}{16 \FPsq}(c_6+2 c_6 g_A^2+7 g_A^2-4 c_4 m_N)\lnmpi 
\nn \\  & & +
q^2\left(2 d_6^{IR} +4 e_{74}^{IR} m_N +\frac{m_N g_A^2}{48 \Fpi^2 
\pi m_\pi} -
\frac{c_4 m_N}{24 \Fpi^2 \pi^2}(1+\lnmpi) \right. \nn \\ & & +
\left. \frac{g_A^2}{48 \FPsq}(7+6 \lnmpi)\right) \, , \\
G_A(q^2)&=&g_A^{IR}+4 d_{16}^{IR} m_\pi^2 -\frac{g_A m_\pi^2}{16
\FPsq}(g_A^2 + (2 g_A^2+1)\lnmpi) \nn \\ & & + \frac{m_\pi^3 g_A}{6
\Fpi^2 \pi}(2 c_4 -c_3)+ \frac{m_\pi^3 g_A}{8 \Fpi^2 \pi
m_N}(g_A^2+1)+q^2 d_{22}^{IR}\, , \label{res:GA} \\
G_P(q^2)&=&\frac{2 \Fpi G_{\pi NN}(m_\pi^2) }{m_\pi^2-q^2}-
2 m_N d_{22}^{IR}  \, \label{res:GP}. 
\eeqa

The correction terms needed to renormalize the LEC's and other
parameters are all proportional to $1/\Fpi^2$. Thus in terms already
containing this factor it doesn't matter which of the various
renormalizations are used as we have consistently neglected terms of
order $1/\Fpi^4$. Thus to simplify the notation in the above, and also
the equations below, we have left off the superscript, $r,IR$, or $EOMS$, on
$g_A$ and the various LEC's when they appear in terms already
containing a $1/\Fpi^2$. Note however that for numerical work we will
always use the appropriate LEC wherever it appears, i.e., when working
in the IR scheme, all LEC's will be the IR values, and similarly for
the other schemes.

Since we kept the isoscalar component of the external vector field we
can also obtain the isoscalar electromagnetic form factors of the
nucleon. (The isovector form factors are of course the same as $G_V$
and $G_M$.)
\beqa
F_1^{(s)}(q^2)&=&1- 4 q^2 d_7^{IR}    \, , \label{res:F1s}\\
F_2^{(s)}(q^2)&=&c_6^{IR}+2c_7^{IR}-32 e_{105}^{IR} m_\pi^2 m_N-
\frac{3 g_A^2 m_\pi^2 }{16 \FPsq}\lnmpi (c_6+2c_7+1) \nn \\ & &
+ 4 q^2(d_7^{IR}+2 e_{54}^{IR} m_N) \,\label{res:F2s} .
\eeqa

Finally the pion-nucleon-nucleon coupling $G_{\pi NN}(q^2)$ is
obtained by identifying the $\pi$NN amplitude of
Eqs.~(\ref{eq:M1pi})-(\ref{eq:M12pi}) with the defining relation
\beq 
- G_{\pi NN}(q^2) \Psibn 
\tdpi\gamma_5 \Psip \, ,   
\eeq 
and is
\beqa
G_{\pi NN}(m_\pi^2) &=&\frac{m_N}{\Fpi}(G_A(0)-2  m_\pi^2 d_{18}^{IR})\,
 .\label{res:GpiNN}
\eeqa

The IR renormalized LEC's, expressed in terms of the EOMS renormalized
LEC's are given by:
\beqa
g_A^{IR}&=&g_A^{EOMS} \label{eq:gaIR} \, , \\
c_6^{IR}&=&c_6^{EOMS}\, ,\label{eq:c6IR} \\
c_7^{IR}&=&c_7^{EOMS}\, , \label{eq:c7IR} \\
d_6^{IR}&=&d_6^{EOMS}+\frac{9 g_A^2}{128 \FPsq}\, , \\
d_7^{IR}&=&d_7^{EOMS}-\frac{3 g_A^2}{256 \FPsq}\, , \\
d_{16}^{IR}&=&d_{16}^{EOMS}-\frac{g_A}{32 \FPsq}(1+g_A^2)+
\frac{c_1 g_A m_N}{16 \FPsq}(4-g_A^2)\, ,\label{eq:d16IR} \\
d_{18}^{IR}&=&d_{18}^{EOMS}-\frac{g_A^3}{192 \FPsq}\, , \\
d_{22}^{IR}&=&d_{22}^{EOMS}+\frac{g_A^3}{192 \FPsq}\, , \\
e_{54}^{IR}&=&e_{54}^{EOMS}+\frac{g_A^2}{512 \FPsq m_N}(1-2 c_6-4 c_7)
\, , \\
e_{74}^{IR}&=&e_{74}^{EOMS}-\frac{g_A^2}{768 \FPsq m_N}(1-2 c_6)\, , \\
e_{105}^{IR}&=&e_{105}^{EOMS}+\frac{3 c_1 g_A^2}{128 \FPsq}(c_6+2 c_7)+
\frac{3 g_A^2}{1024 \FPsq m_N}(4+3 c_6+6 c_7)\, , \label{eq:e105IR}\\
e_{106}^{IR}&=&e_{106}^{EOMS}-\frac{g_A^2}{512 \FPsq m_N}(4-c_6)-
\frac{5 c_1 c_6 g_A^2}{64 \FPsq}\, \label{eq:e106IR}.
\eeqa

The EOMS renormalized LEC's, expressed in terms of the
$\widetilde{MS}$ renormalized LEC's are given by:
\beqa
g_A^{EOMS}&=&g_A^r-\frac{g_A^3 m_N^2}{16 \FPsq}+
\frac{g_A m_N^3}{576 \FPsq}(9 c_2+32 c_3+32 c_4)\,\label{eq:gaEOMS} , \\
c_6^{EOMS}&=&c_6^r+\frac{g_A^2 m_N^2}{16 \FPsq}(c_6+5)\, , \\
c_7^{EOMS}&=&c_7^r-\frac{g_A^2 m_N^2}{16 \FPsq}(4+2 c_6 +3 c_7)\, , \\
d_6^{EOMS}&=&d_6^r+\frac{c_6 g_A^2}{128 \FPsq}\, , \\
d_7^{EOMS}&=&d_7^r-\frac{3 g_A^2}{256 \FPsq}(c_6+2 c_7)\, , \\
d_{16}^{EOMS}&=&d_{16}^r+\frac{c_1 m_N g_A^3}{16 \FPsq}+
\frac{m_N g_A}{288 \FPsq}(c_2+18 c_3-18 c_4-72 c_1) \nn \\ & & +
\frac{c_1 m_N^2 g_A}{1152 \FPsq}(41 c_2+32 c_3+1184 c_4)\, , \\
d_{18}^{EOMS}&=&d_{18}^r-\frac{g_A m_N}{144 \FPsq}(c_2-c_3-c_4)\, , \\
d_{22}^{EOMS}&=&d_{22}^r\, , \\
e_{54}^{EOMS}&=&e_{54}^r+\frac{3 g_A^2}{512 \FPsq m_N}(c_6+2 c_7)\, , \\
e_{74}^{EOMS}&=&e_{74}^r-\frac{c_6 g_A^2}{256 \FPsq m_N}\, , \\
e_{105}^{EOMS}&=&e_{105}^r+\frac{3 c_1 g_A^2}{128 \FPsq}(c_6+2 c_7)\, , \\
e_{106}^{EOMS}&=&e_{106}^r+\frac{3 c_1 c_6 g_A^2}{64 \FPsq}\, .
\eeqa

We find for the $\widetilde{MS}$ renormalized LEC's expressed in terms of
the LEC's of the original Lagrangian:
\beqa
g_A^r&=&g_A + \frac{g_A m_N^2 R}{16 \FPsq}(2-g_A^2)-
\frac{g_A m_N^3 R}{96 \FPsq}(3 c_2 +8 c_3 - 40 c_4) \label{eq:gar} \, , \\
c_6^r&=&c_6 + \frac{g_A^2 m_N^2 R}{32 \FPsq}c_6 \, , \\
c_7^r&=&c_7 - \frac{g_A^2 m_N^2 R}{32 \FPsq}(2 c_6+3 c_7)\, , \\
d_6^r&=&d_6+(1-g_A^2)\frac{R}{192 \FPsq} \, , \\
d_7^r&=&d_7\, , \\
d_{16}^r&=&d_{16}+
\frac{R}{192 \FPsq}\left[ 3 g_A(1-g_A^2)-m_N g_A(c_2+6 c_3-18 c_4)
\right. \nn \\ & & 
\left. - g_A m_N^2 c_1(35 c_2+56 c_3-232 c_4)\right] \, , \\
d_{18}^r&=&d_{18}-\frac{R}{192 \FPsq}g_A m_N(24 c_1+c_2-4c_3-4c_4)\, , \\
d_{22}^r&=&d_{22}\, , \\
e_{54}^r&=&e_{54}\, , \\
e_{74}^r&=&e_{74}+\frac{R}{384 \FPsq m_N}(g_A^2-1-4 c_4 m_N)\, , \\
e_{105}^r&=&e_{105}+
\frac{3 R g_A^2}{1024 \FPsq m_N}(c_6+2c_7)(1+10 c_1 m_N )\, , \\
e_{106}^r&=&e_{106}+
\frac{R}{512 \FPsq m_N}(2 c_6-8 m_N c_4+c_6 g_A^2(1-10 m_N c_1))\, .
\eeqa

Note that the renormalizations of $g_A$ given in Eqs.~(\ref{eq:gaIR}),
(\ref{eq:gaEOMS}), and (\ref{eq:gar}) originate from terms that
survive in the chiral limit and thus they renormalize the original
$g_A$ appearing in the Lagrangian to $g_A$ in the chiral limit,
$\stackrel{\circ}{g}_A$. Just as discussed for the mass in Appendix
\ref{ap:massren}, very often a counterterm to perform this
renormalization is included in the original Lagrangian and $g_A$ is
assumed from the beginning to be $g_A$ in the chiral limit.

\section{Numerical evaluation of LEC's}

In the preceding sections we have obtained the result for the complete
amplitude for OMC as expressed in Eq.~(\ref{OMCamp}) using the values
for the couplings from Eqs.~(\ref{res:GV})-(\ref{res:GP}).  This
amplitude is expressed in terms of the physical masses, the pion decay
constant $\Fpi = 92.4$ MeV, the external parameters $c_1,c_2,c_3,c_4$ and
the sets of LEC's $x^r$, $x^{IR}$, or $x^{EOMS}$, depending on the
case being considered, where $x$ stands for $g_A,
c_6,c_7,d_6,d_7,d_{16},d_{18},d_{22},e_{54},
e_{74},e_{105},e_{106}$. To determine these parameters we use
available data from measurements of weak and electromagnetic form
factors.

For the vector current we have information on the isovector form
factors $F_1^{(v)}$ and $F_2^{(v)}$, equivalent to $G_V$ and $G_M$,
and on the isoscalar form factors $F_1^{(s)}$ and $F_2^{(s)}$. The
static values of the magnetic form factors are given by
$F_2^{(v)}(0)=\kappa_p-\kappa_n$ and $F_2^{(s)}(0)=\kappa_p+\kappa_n$,
where the proton and neutron anomalous magnetic moments are taken as
$\kappa_p= 1.7928$ and $\kappa_n = -1.9130$. We define the slopes of
the various form factors in the usual way
\beq 
F(q^2)=F(0)(1+\frac{q^2}{6}<r^2>)
\eeq
where $q^2$ is the square of the four-vector momentum transfer and
$<r^2>$ is the rms radius. We take the values of the rms radii for
$F_1, F_2$ in the isoscalar and isovector cases from Mergell, {\it et
al.} \cite{Mergell} and thus use $<r^{2(v)}_1>=(0.765 {\rm fm})^2,
<r^{2(v)}_2>= (0.893 {\rm fm})^2,<r^{2(s)}_1>= (0.782 {\rm fm})^2,
<r^{2(s)}_2>= (0.845 {\rm fm})^2$.

Information on the axial current comes from neutron beta decay which
gives $G_A(0)=1.2695 \pm 0.0029$ \cite{PDG} and from
antineutrino-nucleon scattering \cite{Ahrens} which gives the axial
rms radius $<r_A^2>=0.42 \pm 0.04 {\rm fm}^2$. We use for the
pion-nucleon coupling constant $G_{\pi NN}(m_\pi^2) = 13.0 \pm 0.1$
\cite{Stoks}.

There is one remaining unused equation, Eq.~(\ref{res:GP}), which
gives the well known expression for $G_P(q^2)$ in terms of $G_{\pi
NN}(m_\pi^2)$ and $d_{22}$, which can be determined from $<r_A^2>$. 
In principle, if $G_P$ were well
measured, this could be used as an alternative to one of the equations
to determine the LEC's. In view of the uncertainties in the
experimental value of $G_P$ \cite{Gorrev} however this is best used to
predict $G_P$ or simply as a consistency check.

Finally we need the external parameters $c_1,c_2,c_3,c_4$ which can be
obtained from pion nucleon scattering. One should in principle
evaluate these via a complete calculation of pion-nucleon scattering
consistent in order and in its details with the calculation here. That
is beyond the scope of the present paper. So for present purposes we
will simply take the results of a tree level fit obtained by Becher
and Leutwyler \cite{BLpiN}, namely, 
$c_1=-0.9 m_N^{-1}, c_2=2.5 m_N^{-1}, c_3=-4.2
m_N^{-1}, c_4= 2.3 m_N^{-1}$. These parameters appear only in higher
order terms, so this approximation is probably sufficient.

We have identified above nine bits of experimental data to be used to
evaluate the parameters. However there are twelve unknown
parameters. Note however that at least for the IR and EOMS schemes at
leading order only certain combinations appear. Thus we define
\beqa \label{eq:gpLEC}
\tilde{g}_A^{IR}&=&g_A^{IR} +  4 m_\pi^2 d_{16}^{IR}\,  , \nn  \\
\tilde{c}_6^{IR}&=&c_6^{IR} - 16 m_\pi^2 m_N e_{106}^{IR} \,  , \nn \\
\tilde{c}_7^{IR}&=&c_7^{IR} -  8 m_\pi^2 m_N(2e_{105}^{IR}- e_{106}^{IR})\,  ,
\eeqa
with an analogous definition for the EOMS and $\widetilde{MS}$
schemes. For the IR and EOMS schemes, which obey counting, the
$m_\pi^2$ coefficient in these definitions means that we can replace
all $g_A, c_6, c_7$ appearing in higher order terms with $\tilde{g}_A,
\tilde{c}_6, \tilde{c}_7$. Thus we eliminate all instances of
$d_{16},e_{105},$ and $e_{106}$, and so have enough input data to
solve uniquely for the nine parameters.

For the $\widetilde{MS}$ scheme however this doesn't work. Because of
the non counting terms the replacement $g_A, c_6, c_7 \rightarrow
\tilde{g}_A, \tilde{c}_6, \tilde{c}_7$ leaves some instances of
$d_{16},e_{105},$ and $e_{106}$ which are not of higher order. Thus we
need to assign values to these LEC's in order to solve for the
others. Since there is not enough experimental information available
we will simply try a couple of arbitrary cases to get a feel for the
sensitivity of the results to these LEC's. In particular we will take,
for a case $\widetilde{MS}$-a, $d_{16}=e_{105}=e_{106}=0$. As an
alternative we will take for case $\widetilde{MS}$-b,
$d_{16}=e_{105}=e_{106}=1$, expressed in appropriate units. This
latter choice is arbitrary, but should correspond to a 'natural' size
for these LEC's.

To actually solve for the LEC's for say the IR case we take
Eqs.~(\ref{res:GV})-(\ref{res:GA}),(\ref{res:F1s}),(\ref{res:F2s}) and
(\ref{res:GpiNN}) and express all of the LEC's in terms of their IR
forms, so that the equations are expressed purely in terms of IR
quantities. We then solve these equations self consistently, using the
experimental input given above, for all the LEC's. In particular this
means that we solve the cubic equation for $\tilde{g}_A^{IR}$ and use
that value in the other equations to solve for the other LEC's. To get
the EOMS case we use Eqs.~(\ref{eq:gaIR})-(\ref{eq:e106IR}) to replace
the IR LEC's with their EOMS forms, dropping higher order terms as
appropriate, so that the equations are given entirely in terms of EOMS
quantities, and then repeat the solution procedure. Note that this
procedure corresponds to what one would do if one were using the EOMS
scheme from the beginning. It is not quite the same as simply using
Eqs.~(\ref{eq:gaIR})-(\ref{eq:e106IR}) to get the EOMS LEC's from the
IR results because of the numerical consequences of higher order terms
which would be treated slightly differently in those two
approaches. Finally one gets the $\widetilde{MS}$ results in analogous
way, though here as noted above, for that case we have to choose
values for $d_{16},e_{105},$ and $e_{106}$.

The results for the LEC's obtained as described above by consistently
solving all the relations available from the OMC amplitude are given
in Table~\ref{LECresults}, together with available results obtained by
others.

\begin{table}
\caption{Results for $\tilde{g}_A$ and the various LEC's in each of
the renormalization schemes discussed in this work. Given for
comparison are results from \protect{\cite{km-npa01}} and
\protect{\cite{fgs-jpg04}} converted to account for the normalization,
to $m_N$ vs $m$, which we use, and in the case of
Refs.~\protect{\cite{fgs-jpg04},\cite{Fuchsthesis}} for the different
combinations of $\tilde{c}_6$ and $\tilde{c}_7$ they use. The parameters
$\tilde{g}_A$, $\tilde{c}_6$, and $\tilde{c}_7$  are dimensionless, 
and the $d_i$ and $e_i$ have
respectively units of GeV$^{-2}$ and GeV$^{-3}$. The cases labeled
$\widetilde{MS}$-a and $\widetilde{MS}$-b involve arbitrary choices of
$d_{16},e_{105},$ and $e_{106}$ as described in the text. }
\label{LECresults}
\begin{tabular}{c|c|ccccccccccccc}
&&$\tilde{g_A}$ & $\tilde{c_6}$ & $\tilde{c_7}$ & $d_6$ & $d_7$ & $d_{18}$ &
$d_{22}$ & $e_{54}$ & $e_{74}$ & $d_{16}$ & $e_{105}$ & $e_{106}$ \\
\hline
      & IR & 0.9568 & 4.45 & -2.34 & 0.07 & -0.65 & -0.25 & 
2.28 & 0.30 & 2.16 &- &- &-  \\
This  & EOMS & 1.1030 &6.35 & -3.26 & -0.57 & -0.49 & -0.17& 
2.20 & 0.26 & 1.62 &-&- &- \\
work   & $\widetilde{MS}$-a &-0.6244  &2.52&-0.49&-1.01&-0.52&
-0.49&2.30&0.28&2.72&0.0&0.0&0.0 \\
& $\widetilde{MS}$-b &-0.5810  &2.29&0.66&-1.01&-0.44&
-0.48&2.29&0.26&2.77&1.0&1.0&1.0 \\
\hline
Ref.~\protect{\cite{km-npa01}} & IR &1.26 &5.18&-2.77&0.80&
-0.75&-&-&0.26&1.65&-&-&- \\
\hline
Ref.~\protect{\cite{fgs-jpg04},\cite{Fuchsthesis}} & IR &1.267 &
4.73&-2.54&0.59&-0.79&-&-&0.25&1.93&-&-&- \\
 & EOMS &1.267 &4.73&-2.49&-0.75&-0.54&-&-&0.19&1.59&-&-&- \\
\end{tabular}

\end{table}

First we should comment on the comparison with previous results. There
have been two previous calculations of the electromagnetic form
factors and the corresponding LEC's in relativistic formulations,
Refs.~\cite{km-npa01},\cite{fgs-jpg04}. While our results are
qualitatively the same, there are differences in detail.

Perhaps the main difference in principle is the value of $g_A$
used. In both of these previous works $g_A$ was taken to be $g_A
\simeq G_A(0)=1.26$ which is the lowest order result of
Eq.~(\ref{res:GA}). Also since $d_{16}$ doesn't appear explicitly in
the vector current it was not necessary there to distinguish between
$g_A$ and $\tilde{g}_A$.  We however expressed everything in terms of
$\tilde{g}_A$ and solved Eq.~(\ref{res:GA}) consistently to the order
of the calculation to obtain a value of $\tilde{g}_A$. Since $g_A$
appears in many places, and in particular in the corrections to all
the other LEC's, this made a difference, significant in some cases, in
the values of the LEC's obtained. In a purely formal sense the
corrections to $g_A$, i.~e. the differences between $g_A$,
$\tilde{g}_A$, and the lowest order approximation 1.26 are all of
higher order. Thus in principle the use of any of these three
interchangeably in the formulas for the LEC's would be consistent with
our other approximations. The fact that it makes a difference simply
reflects the fact the the higher order terms are not always small,
i.~e. that the expansion doesn't always converge well. However since
we have the information, via Eq.~(\ref{res:GA}), to calculate the
corrections to $g_A$, it seems appropriate and preferable to use that
information consistently in obtaining the other LEC's. Finally note
that some further differences arise because in our self consistent
solution for $\tilde{g}_A$ the results are different for the IR and
EOMS cases, because $d_{16}$ is different for those cases.

Additional smaller differences arise because we used the rms radii
appropriate to the Dirac and Pauli form factors $F_1, F_2$ which were
the form factors calculated directly, rather than converting to radii
appropriate for the Sachs form factors. This affects some of higher
order terms and seems to affect $d_6$ particularly. Also, as discussed
above, there are different options for including mass insertions and
for expanding to get the non counting terms and we did not always use
the same conventions as in previous work. The value of $c_4$ used was
slightly different than the one in Ref.~\cite{km-npa01}. Finally we
expressed everything in terms of the physical mass $m_N$ instead of
$m$. Again formally these should be interchangeable, but numerically
it made a difference in some cases.

While we think the use of the self consistent value of $\tilde{g}_A$
is a definite improvement in principle over previous works, the other
differences are really just differences in the details of the
calculation. The fact that they make a numerical difference in the
values of the LEC's just reinforces the statement made at the
beginning. Namely, if one wants values of the LEC's which can be used
in further calculations one must be sure that the same approach and
the same conventions and approximations are made. Otherwise it is
dangerous to simply lift results from one calculation to use in
another.
  
Now let's look more carefully at the results of
Table~\ref{LECresults}. Note that $\tilde{g}_A, \tilde{c}_6,
\tilde{c}_7$ differ for the IR and EOMS cases. Since the underlying
parameters $g_A, c_6, c_7$ don't change (cf. Eqs.~(\ref{eq:gaIR}),
(\ref{eq:c6IR}), and (\ref{eq:c7IR})) these differences must be due
to differences in $d_{16}, e_{105}, e_{106}$ as given in
Eqs.~(\ref{eq:d16IR}), (\ref{eq:e105IR}), and (\ref{eq:e106IR}). The LEC
$d_6$ is very different for IR and EOMS schemes, and also varies from
previous results.  this is apparently because of strong cancellations
among terms, which make it very sensitive to the small
corrections. Our value of $\tilde{c}_6$ for the EOMS case differs
significantly from that of Ref.~\cite{fgs-jpg04} apparently because of
the $c_1 c_6$ and $c_1 c_7$ terms we have (cf. Eqs.~(\ref{eq:e105IR})
and (\ref{eq:e106IR})) which they have not kept. These terms seem to
originate in the different way of including the mass insertions which
we used.

In the $\widetilde{MS}$ scheme the parameters $\tilde{g}_A,
\tilde{c}_6, \tilde{c}_7$, and $d_6$ change fairly dramatically as
compared with values obtained in the IR or EOMS schemes. Apparently
$d_6$ is still sensitive to cancellations and the other three contain
large non counting terms, which also do not vanish in the chiral
limit. Had we adopted the common procedure of first adding a
counterterm to the Lagrangian to renormalize $g_A$ to the chiral
limit, such large terms would not be there for $g_A$, and presumedly
it would be the same as for the IR and EOMS schemes. However such
large terms would still be present for $c_6$ and $c_7$ and would still
affect $\tilde{g}_A, \tilde{c}_6, \tilde{c}_7$ via the values of
$d_{16}, e_{105}, e_{106}$ buried in them.

Note that all the results for the $\widetilde{MS}$ scheme are
dependent on the somewhat arbitrary choices made for $d_{16}, e_{105},
e_{106}$. The two illustrative cases correspond to values of zero for
these LEC's and values of unity in natural units. Many of the LEC's
are similar for the two cases but a few, particularly $\tilde{c}_7$,
change a lot. Clearly if one wants to seriously use the
$\widetilde{MS}$ scheme, it will be necessary to pin down $d_{16},
e_{105}, e_{106}$ from some other process.

Finally we should make a few general remarks. All three of these
schemes, since they differ only in how they absorb or do not absorb
the finite non counting terms in the LEC's will give the same values
for the amplitudes. One might hope that one scheme or another would,
say, lead to all small LEC's which could be neglected. This does not
seem to be the case and there does not seem to be any general pattern
emerging when we compare the three schemes. The parameters
$\tilde{g}_A, \tilde{c}_6, \tilde{c}_7$ are perhaps a bit smaller in
the $\widetilde{MS}$ scheme than in the others, indicating that the
specific non counting terms which are kept explicit in the
$\widetilde{MS}$ scheme but absorbed in the LEC's in the other schemes
are large. However this does not persist for the $d$'s or $e$'s which
are of the same size, or maybe smaller, in the IR and EOMS schemes as
in the $\widetilde{MS}$ scheme.

\section{Discussion of different approaches}

In previous sections we have described an explicit calculation - that
of the amplitude for OMC - carried out in three different schemes for
Lorentz invariant chiral perturbation theory. In this section we want
to compare and contrast these schemes, particularly from the point of
view of how best to do a practical calculation.

First, as a matter of principle the IR and EOMS schemes are major 
advances in
our understanding of how to handle Lorentz invariant ChPT calculations. 
Such approaches show that in general it is possible to rewrite
relativistic ChPT so that it obeys the same counting rules as HBChPT,
which thus solves the problem with such theories raised in
\cite{Gasser}. It was also shown, particularly in \cite{BL} that,
unlike HBChPT, these schemes preserved the correct analytic structure
of the amplitudes. That feature has not been important for the OMC
calculation, but can be for other processes.

Thus we now know that in a relativistic theory the choice of number of
loops and the choice of the order of the Lagrangian to use at each
vertex can be made in a rigorous way that preserves HBChPT counting,
and that low order contributions from higher order diagrams can all be
absorbed in a consistent way in the LEC's. From the point of view of a
practical calculation that means that the choice of diagrams and
vertices can be made essentially as in HBChPT.

Once that choice is made however, from a practical point of view, one
has options. We have considered three possibilities for
renormalization: IR, EOMS and $\widetilde{MS}$. All three treat the
infinities, i.~e. the $R$ terms in the same way. They differ only in
which subset of the set of finite terms which do not obey
counting are absorbed in the LEC's. Thus the LEC's will have different
numerical values in the three schemes and the formulas for measurable
quantities will look different. But all three will give the same
predictions for measurable results. Once the general principles have
been used to choose the diagrams to be considered, any one of the
three schemes could be used consistently for practical calculations
and would give equivalent results.

We can discuss however some of the pros and cons of the three schemes,
relative to practical calculations.

Consider first the IR approach. It absorbs the largest number of terms
in the LEC's and as a consequence the formulas tend to look
simpler. However one might be hiding known physics by absorbing such
terms. This approach is probably the simplest of the three as long as
one does not need to work out the exact formulas for renormalization
of each of the LEC's. This is because if one just 'drops' the terms
which would later be absorbed one can drop a lot of integrals - all
with only nucleon propagators - and thus reduce the number of diagrams
to be calculated. If one calculates explicit formulas for the
renormalization of the LEC's, which we have done here, though it would
not normally be really necessary, then all diagrams have to be
calculated for all three schemes.

In contrast the $\widetilde{MS}$ scheme absorbs none of the finite
terms. It is thus closest to the historical approach of describing a
process by a set of Feynman diagrams. Some non counting terms will
appear, but may be considered to have physical significance. An
example of this can be seen in the classical approaches to radiative
corrections to neutron beta decay where certain terms, which in the
relativistic ChPT approach seem to originate as non counting
contributions from diagrams too high order to keep \cite{aetal-plb04},
appear explicitly in the standard Feynman diagram approach
\cite{Sirlin}, have been discussed individually \cite{MarcSir}, and
are considered relevant.

The $\widetilde{MS}$ scheme requires more effort than the IR scheme,
if explicit formulas for the renormalization are not required, as one
must always calculate all diagrams. Since there are non counting terms
still present, the grouping of LEC's to reduce the number of
independent quantities to be fitted to experiment, as done in
Eq.~(\ref{eq:gpLEC}), will not necessarily work, as we saw for the
present calculation. This is a serious disadvantage for a single
calculation as it increases the number of LEC's to be evaluated from
data. It might be less of a problem for a series of calculations as it
is unlikely that the same grouping will work for all processes and so
in that case for all the schemes one probably has to evaluate all
LEC's individually anyway.

The EOMS scheme is somewhere in between the other two. It absorbs the
minimum number of terms necessary to get counting. It thus may preserve
some of the good things about the $\widetilde{MS}$ scheme while still
solving the counting problem. It however requires the most work of all
as every diagram must be evaluated and then one must look at each
diagram individually to determine which terms to subtract. It also
requires a careful statement of conventions, as discussed above.

In a general sense the LEC's absorb our ignorance, so it would seem
that one would want to leave explicit as much known physics as
possible, and absorb as little as possible into the LEC's. Ideally the
LEC's representing unknown physics would then get small. This is
the general philosophy behind attempts to include explicitly
additional degrees of freedom, such as the $\Delta$ \cite{hacker,
bern03} or vector mesons \cite{schlin05,km-npa01}.  Thus smallness of
the LEC's might be a criterion for the choice of scheme. One has no
knowledge of the size or sign of the sum of terms contributing to an
LEC from higher order diagrams however. Also, in the present example,
OMC, there is no obvious choice leading to small LEC's, so it is not
clear how to implement this criterion.

\section{Conclusions}

We can thus summarize as follows. We have evaluated the OMC amplitude
through ${\cal O}(p^4)$ in the three schemes, $\widetilde{MS}$, EOMS
and IR. Using available data we have solved self consistently for the
nine LEC's which appear in the IR or EOMS schemes. The
$\widetilde{MS}$ scheme requires three additional LEC's, for which
further data would be required. Similar evaluations of the LEC's for
the vector current have been done before, and our results differ from
these primarily because we have self consistently solved the equations
coming from the axial current for $g_A$ and have used that value,
rather than the lowest order result used in previous work. Many
subtleties and details of the calculation also affect the numerical
values of the LEC's, which indicates that before using these or other
values of the LEC's to calculate new processes it will always be
necessary to make sure that the new calculation is done in exactly the
same way as that used to extract the LEC's.

\section*{Acknowledgments}

The authors would like to thank M. Schindler for useful conversations
and S. Scherer both for a careful reading of the manuscript and for
some very useful comments. Figures were prepared using the program
JaxoDraw \cite{jaxodraw} provided by L.~Theussl.  This work was
supported in part by the Natural Sciences and Engineering Research
Council of Canada.  SA is supported in part by the Korean Research
Foundation and a Korean Federation of Science and Technology Societies
Grant funded by Korean Government (MOEHRD, Basic Research Promotion
Fund). HWF would like to acknowledge the hospitality of the Aspen
Center for Physics where part of this work was done. He also would
like to thank Prof. Jim Sheppard of the University of Colorado for his
hospitality.

\appendix
\section{Loop integrals}
\label{ap:loopinteg}

We will define the general loop integral in $d$ dimensions containing
$i$ pion propagators and $j$ nucleon propagators and corresponding to
the momenta as in Fig.~\ref{fig:genloop} as
\beq
I_{\pi\pi...\pi N N...N}[k_1,k_2,...,k_i,p_1,p_2,...p_j,A]=
i \mu^{4-d}\int \frac{d^d\ell}{(2 \pi)^d}\frac{A}
{D_\pi(k_1)...D_\pi(k_i)D_N(p_1)...D_N(p_j)} \, .
\eeq
Here $\mu$ is a scale factor and $A$ is the numerator function, which
may contain anything. $D_\pi(k)=(\ell+k)^2-m_{0\pi}^2+i \epsilon$ and
$D_N(p)=(\ell+p)^2-m^2+i \epsilon$ are respectively the pion and
nucleon propagator denominators. $m_{0\pi}$ and $m$ are the
unrenormalized pion and nucleon masses appearing in the original
Lagrangian. Thus the number of subscripts $\pi$ and $N$ correspond to
the number of pion and nucleon propagators respectively. We will
always redefine the integration variable $\ell$ so as to
make the first pion momentum $k_1=0$ and will drop it from the
argument list.

In general we can always reduce $A$ to factors which can be removed
from the integral or to powers of $\ell^\mu$, which at the one loop
level can be reduced out using standard tensor expansions. Thus the
only integrals which need to be evaluated explicitly have $A=1$.

For the basic calculation we need the integrals $I_{\pi}[1]$,
$I_{\pi\pi}[q, 1]$, $I_{N}[p,1]$, $I_{NN}[p_i,p_f,1]$, $I_{\pi
N}[p,1]$, $I_{\pi NN}[p_i,p_f,1]$, $I_{\pi \pi N}[-q,p_i,1]$. For
those diagrams with a mass insertion on an internal nucleon diagram,
which duplicates one of the nucleon propagators, we require the
additional integrals $I_{NN}[p,p,1]$, $I_{NNN}[p_i,p_f,p_f,1]$,
$I_{NNN}[p_i,p_i,p_f,1]$, $I_{\pi NN}[p,p,1]$, $I_{\pi
NNN}[p_i,p_f,p_f,1]$, $I_{\pi \pi NN}[-q,p_i,p_i,1]$, where $p$
can be either $p_i$ or $p_f$ and where $q=p_f-p_i$.
  
The evaluation of these integrals in the form needed for the IR or
EOMS schemes proceeds in the standard fashion, as described for
example in \cite{BL}. The meson and nucleon propagators are separately
combined using the Feynman parameter approach. The two pieces are then
combined and the infinities extracted using standard dimensional
regularization formalism. The results can then be expressed in $d$
dimensions in terms of the $R$ and $\epsilon=(4-d)/2$ of
Eq.~(\ref{Rdef}) and of relatively simple integrals over the Feynman
parameters. This approach however leads, as discussed in the main
text, to results which do not obey the usual HBChPT counting rules.

Becher and Leutwyler \cite{BL} modify this procedure by dividing the
integrals into two parts, one containing the infrared singularities
and the other a regular polynomial in the expansion parameter. The
regular part is then 'dropped', i.e. in a formal sense absorbed in the
LEC counterterms.

In order to discuss both the standard and the Becher-Leutwyler
approach simultaneously we define a parameter $\BL$ which
flags the terms to be dropped in the Becher-Leutwyler procedure. Thus
integrals involving only nucleons obtain an overall $\BL$ in
accord with the result that they are regular. Integrals with only pions
are evaluated in the standard approach and so contain no
$\BL$. For those integrals involving both pions and nucleon
propagators the basic integral on the parameter $z$ is divided into two
parts and evaluated in accord with \cite{BL,Sch04} as
\beq
\int_{0}^{1} dz \rightarrow \int_{0}^{\infty} dz \, - \,  
\BL \int_{1}^{\infty} dz \, .
\eeq
As discussed above, the singular terms proportional to $R$, which
appear in both regular and infrared parts, can be recombined (i.e we
eventually put $\BL R \rightarrow R$) and the renormalization of the
LEC's carried out in the usual way. Thus $\BL$ will serve to flag the
regular terms which would be dropped in the
Becher-Leutwyler procedure.

With these preliminaries recorded we can list the results for the
integrals we need in this calculation.

$I_{\pi}[1]$, $I_{\pi\pi}[q, 1]$, $I_{N}[p,1]$, and $I_{NN}[p_i,p_f,1]$
are standard and the results are given here for completeness only:

For $I_{\pi}[1]$
\beq
I_{\pi}[1] = \frac{m_{0\pi}^2}{(4 \pi)^2}(R+\lnmpib) \, .
\eeq
For $I_{\pi\pi}[q, 1]$
\beq
I_{\pi\pi}[q, 1] = \frac{1}{(4 \pi)^2}(R+1+\lnmpib)+W_{\pi\pi}[q^2] \, ,
\eeq
with
\beqa
W_{\pi\pi}[q^2]&=&\frac{1}{(4\pi)^2}\int_{0}^{1} dz \: {\rm ln}(D-i \eta)\\
&=& - \frac{q^2}{96 \pi^2 m_{0\pi}^2}-\frac{q^4}{960 \pi^2 m_{0\pi}^4}+ ....
\eeqa
where here
\beq
D=1-\frac{q^2}{m_{0\pi}^2}z(1-z) \, .  
\eeq
For $I_{N}[p,1]$
\beq
I_{N}[p,1] = I_{N}[0,1] = \BL \frac{m^2}{(4 \pi)^2}(R+\Logm) \, .
\eeq
For $I_{NN}[p_i,p_f,1]$
\beqa
I_{NN}[p_i,p_f,1] &=&\BL\frac{1}{(4 \pi)^2}(R+1+\Logm)+W_{NN}[q^2]\, , \\
I_{NN}[p,p,1] &=& I_{NN}[0,0,1]\ =\BL\frac{1}{(4 \pi)^2}(R+1+\Logm)\, , \\
I_{NNN}[p_i,p_f,p_f,1] &=&I_{NNN}[p_i,p_i,p_f,1] = W_{NNN}[q^2] \, , 
\eeqa
with
\beqa
W_{NN}[q^2]&=&\BL \frac{1}{(4\pi)^2}\int_{0}^{1} dz \: {\rm ln}(D-i \eta)\\
&=& \BL(-\frac{q^2}{96 \pi^2 m_N^2}+
\frac{\dm q^2 m_{0\pi}^2}{96 \pi^2 m_N^4}-
\frac{q^4}{960 \pi^2 m_N^4}+ ....)\, , \\
W_{NNN}[q^2]&=&\frac{\BL}{(4 \pi)^2 m^2}\int_{0}^{1} \frac{dz}{D-i\eta}\\
&=& \BL(\frac{1}{32 \pi^2 m_N^2}-
\frac{\dm m_{0\pi}^2}{32 \pi^2 m_N^4}+
\frac{q^2}{192\pi^2 m_N^4}+....)
\eeqa
where here
\beq
D=1-\frac{q^2}{m^2}z(1-z) \quad {\rm and} 
\quad \delta_m=\frac{m^2-m_N^2}{m_{0\pi}^2} \, .
\eeq

For simplicity we have given only the first few terms in the
expansions of the $W$'s above, as the full expressions used are quite
lengthy.  In the actual calculations we kept more terms, as many as
necessary to obtain the final amplitude through the first four orders
in the expansion parameter.

The remaining integrals involve both pion and nucleon propagators. For
those we follow and generalize the procedure used in \cite{BL} for
$I_{\pi N}$.

For $I_{\pi N}[p,1]$ we find, after combining denominators and
evaluating via dimensional regularization,
\beq
I_{\pi N}[p,1] = -\frac{1}{(4 \pi)^2}(R-1+
\Logm)(2 \epsilon -1)\int_{0}^{1}\frac{dz}{(C-i\eta)^\epsilon} \, .
\eeq
Here $C=C_0+C_1 (z-z_0)^2$. For this case $C_1=1+2 \alpha \Omega
+\alpha^2$, $C_0=\alpha^2(1-\Omega^2)/C_1$ and $z_0=\alpha
(\alpha+\Omega)/C_1$. Here (and below) $\alpha=m_{0\pi}/m$ and
$\Omega=(p^2-m^2-m_{0\pi}^2)/(2 m m_{0\pi})$.  These integrals depend on
the square of the four momentum $p^2$ but we will need them only at
the physical on shell point $p^2=m_N^2$. We must account for the fact
that $m \neq m_N$ and hence as above use $m^2-m_N^2 \equiv \dm
m_{0\pi}^2$ where $\dm$ is a dimensionless parameter presumedly
of order one. In fact, from Appendix~\ref{ap:massren} we have $\dm = 8 c_1
m_N+...$.  This allows us to expand in powers of $m_{0\pi}$ about the
physical mass $m_N$.
The integral can be done analytically, basically by integrating by
parts, as in \cite{BL} and we obtain
\beq
I_{\pi N}[p,1]=\frac{R}{(4\pi)^2}(\BL-\frac{m_{0\pi}^2}{2 m_N^2}
(\BL-1)(1-\dm))+
W_{\pi N}[p^2,m^2,m_{0\pi}^2] \, ,
\eeq
where on shell
\beqa
W_{\pi N}[m_N^2,m^2,m_{0\pi}^2]&=& \frac{\BL}{16 \pi^2}(\LogmN-1)+
\frac{m_{0\pi}}{16 \pi m_N} +  
\frac{m_{0\pi}^2}{32 \pi^2 m_N^2}(1-
3\BL  \nn \\ & & +\dm(\BL-1)+(1-\dm)(\lnmpib-\BL\LogmN))+.... \, .
\eeqa

In a similar fashion we find for $I_{\pi NN}[p_i,p_f,1]$,$I_{\pi
NNN}[p_i,p_f,p_f,1]$, and $I_{\pi NNN}[p_i,p_i,p_f,1]$
\beqa
I_{\pi NN}[p_i,p_f,1]&=&-\frac{1}{2 m^2 (4 \pi)^2}(R+1+\Logm)(2 
\epsilon)\int_{0}^{1} dy \int_{0}^{1}\frac{z \: dz}
{(C-i\eta)^{1+\epsilon}}\, , \\ 
I_{\pi NNN}[p_i,p_f,p_f,1]&=&I_{\pi NNN}[p_i,p_i,p_f,1]  \nn \\
&=&-\frac{1}{ m^4 (4 \pi)^2}(1+2 \epsilon)
\int_{0}^{1} y dy  \int_{0}^{1}\frac{z^2 dz}
{(C-i\eta)^{2+\epsilon}}  \, ,
\eeqa
where now on shell with $p_i^2=p_f^2=m_N^2$, $C_1=1+2 \alpha \Omega
+\alpha^2-y(1-y)q^2/m^2$, $C_0=\alpha^2(1-\Omega^2-y(1-y)q^2/m^2)/C_1$
and $z_0=\alpha (\alpha+\Omega)/C_1$.
The integrals on $z$ can be obtained analytically by generalizing the
procedure of \cite{BL}. This result is then expanded in powers of the
small parameter and integrated term by term on $y$. We thus obtain 
\beqa
I_{\pi NN}[p_i,p_f,1]&=&W_{\pi NN}[p_i^2,p_f^2,q^2,m^2,m_{0\pi}^2]\, , \\
I_{\pi NNN}[p_i,p_f,p_f,1]&=&I_{\pi NNN}[p_i,p_i,p_f,1]=\frac{1}
{m_{0\pi}}W_{\pi NNN}[p_i^2,p_f^2,q^2,m^2,m_{0\pi}^2] \, ,
\eeqa
where on shell
\beqa
W_{\pi NN}[m_N^2,m_N^2,q^2,m^2,m_{0\pi}^2]&=&\frac{\BL-1}{32 \pi^2 m_N^2}+
\frac{1}{32 \pi^2 m_N^2}(\BL \LogmN-\lnmpi)  \nn \\&+&
\frac{m_{0\pi}}{64 \pi m_N^3}(1-\dm)+
\frac{m_{0\pi}^2}{64 \pi^2 m_N^4}(1-2\BL)(1-2 \dm) \nn \\ &+&
\frac{m_{0\pi}^2 \dm^2}{64\pi^2 m_N^4} -
\frac{q^2}{192 \pi^2 m_N^4}(1+\lnmpi \nn \\ &-& \BL \LogmN)+ .....\, , \\
W_{\pi NNN}[m_N^2,m_N^2,q^2,m^2,m_{0\pi}^2]&=&-\frac{1}{128 \pi m_N^3}-
\frac{m_{0\pi}}{64 \pi^2 m_N^4}(1-2\BL-\dm)  \nn \\&-&
\frac{3 m_{0\pi}^2}{1024 \pi m_N^5}(1-6\dm+\dm^2)-
\frac{q^2}{512 \pi m_N^5}+... \, .
\eeqa
The $\pi NN$ integral with duplicate nucleon propagators can be
obtained by taking $q^2 \rightarrow 0$ in $I_{\pi NN}[p_i,p_f,1]$,
namely
\beq
I_{\pi NN}[p,p,1] =W_{\pi NN}[m_N^2,m_N^2,0,m^2,m_{0\pi}^2] \, . 
\eeq

Finally
\beqa
I_{\pi \pi N}[-q,p_i,1]&=&-\frac{1}{2 m^2 (4 \pi)^2}(R+1+
\Logm)(2 \epsilon)\int_{0}^{1} dx \int_{0}^{1}\frac{(1-z) dz}{(C-
i\eta)^{1+\epsilon}} \, , \\
I_{\pi \pi NN}[-q,p_i,p_i,1]&=&-\frac{1}{ m^4 (4 \pi)^2}(1+2 \epsilon)
\int_{0}^{1} dx  \int_{0}^{1}\frac{(1-z)z dz}{(C-i\eta)^{2+\epsilon}}  \, ,
\eeqa
where we now have on shell with $p_i^2=m_N^2$, $C_1=1+2 \alpha \Omega
+\alpha^2-x(1-x)q^2/m^2$, $C_0=\alpha^2-C_1 z_0^2-x(1-x)q^2/m^2$ and
$z_0=\alpha (\alpha+\Omega)/C_1-x(1-x)q^2/( m^2 C_1)$.
\cite{footnote3}
Again we can do the $z$ integration analytically and then expand in
powers of the small parameter and do the $x$ integration term by
term. We then obtain
\beqa
I_{\pi \pi N}[-q,p_i,1]&=&
\frac{1}{m_{0\pi}}W_{\pi \pi N}[p_i^2,p_f^2,q^2,m^2,m_{0\pi}^2]\, , \\
I_{\pi \pi NN}[-q,p_i,p_i,1]&=&
\frac{1}{m_{0\pi}^2}W_{\pi\pi NN}[p_i^2,p_f^2,q^2,m^2,m_{0\pi}^2] \, ,
\eeqa
where on shell
\beqa
W_{\pi \pi N}[m_N^2,m_N^2,q^2,m^2,m_{0\pi}^2]&=&
\frac{1}{32 \pi m_N}(1+\frac{q^2}{12 m_{0\pi}^2})  \nn \\ &+&
\frac{m_{0\pi}}{32 \pi^2 m_N^2}(2-\dm-3\BL+
\lnmpi-\BL \LogmN)  \nn \\ &-&
\frac{q^2}{192 \pi^2 m_N^2 m_{0\pi}}(1+\dm)+ .....\, , \\
W_{\pi\pi NN}[m_N^2,m_N^2,q^2,m^2,m_{0\pi}^2]&=&-
\frac{1}{32 \pi^2 m_N^2}(1+\frac{q^2}{6 m_{0\pi}^2})+
\frac{m_{0\pi}}{128 \pi m_N^3}(1+\dm) \nn \\ &+&
\frac{q^2}{1536 \pi m_N^3 m_{0\pi}}(1+3\dm)+.... \, .
\eeqa

Finally we observe that there is an alternative method for obtaining
the loop integrals involving two nucleon propagators of the same
momentum. It follows from the relations
\beq
\frac{\partial}{\partial m^2} \{ \frac{1}{p^2-m^2 }\}=
 \{ \frac{1}{p^2-m^2} \}^2, \quad \quad \frac{\partial}{\partial m^2}=
\frac{1}{m_{0\pi}^2}\frac{\partial}{\partial \dm}
\eeq
 that one can get a loop integral with a
duplicate propagator by taking derivatives, namely
\beqa
\frac{1}{m_{0\pi}^2} \frac{\partial}{\partial \dm}I_{\pi N}[p,1]&=&
I_{\pi NN}[p,p,1]\, , \\
\frac{1}{m_{0\pi}^2} \frac{\partial}{\partial \dm}I_{NN}[p_i,p_f,1]&=& 
I_{NNN}[p_i,p_f,p_f,1]+I_{NNN}[p_i,p_i,p_f,1]\, , \\
\frac{1}{m_{0\pi}^2} \frac{\partial}{\partial \dm}I_{\pi NN}[p_i,p_f,1]&=&
I_{\pi NNN}[p_i,p_f,p_f,1]+I_{\pi NNN}[p_i,p_i,p_f,1]\, , \\
\frac{1}{m_{0\pi}^2} \frac{\partial}{\partial \dm} I_{\pi \pi N}[-q,p_i,1]
&=& I_{\pi \pi NN}[-q,p_i,p_i,1] \, .
\eeqa

We have checked that our results satisfy these relations. Note that to
obtain $I_{\pi NN}[p,p,1]$ for example to ${\cal O}(m_\pi^4)$ we need
$I_{\pi N}[p,1]$ to ${\cal O}(m_\pi^6)$ because of the $m_\pi^2$
introduced in the denominator by the derivative. This somewhat lessens
the utility of this method for actually calculating the integrals with
duplicate propagators.

\section{Mass and wave function renormalization}
\label{ap:massren}

In the meson sector the pion mass and wave function renormalizations
are calculated in standard fashion and are given in a number of
sources, for example, \cite{fear97}. In our conventions and notation
we have
\beq \label{physpimass}
   m_\pi^2 = m_{0\pi}^2\left[1+\frac{2m_\pi^2}{\Fpi^2}\left(l_3^r(\mu)+
             \frac{1}{4(4\pi)^2}{\rm ln}\left(\frac{m_\pi^2}{\mu^2}\right)
             \right)\right],
\eeq
and 
\beq \label{defZpi}
   Z_\pi
      = 1 - \frac{2m_\pi^2}{\Fpi^2}\left[l_4^r(\mu)+\frac{2}{3(4\pi)^2}R
      -\frac{1}{3(4\pi)^2}{\rm ln}\left(\frac{m_\pi^2}{\mu^2}\right)\right]~.
\eeq
where the LEC's have been renormalized as 
\beqa
 l_3^r(\mu) &=& l_3+\frac{R}{4(4\pi)^2} \, , \\
 l_4^r(\mu) &=& l_4-\frac{R}{(4\pi)^2}\, .
\eeqa

The renormalization of the pion decay constant is also standard and
given from \cite{fear97} by
\beq
\Fpi = \F \left[1+\frac{m_\pi^2}{\Fpi^2}\left(l_4^r(\mu)-\frac{1}{(4\pi)^2}
                  {\rm ln}\left(\frac{m_\pi^2}{\mu^2}\right)\right)\right]~.
\eeq

In the pion-nucleon sector the nucleon mass and wave function
renormalizations must be calculated in a fashion consistent with the
rest of the calculation. The appropriate diagrams contributing to the
nucleon self energy are given in Fig.~\ref{fig:nucselfen} and the
amplitudes corresponding to those figures are given by
\beqa
M_{1NN} &=& i \Psib (\sla{p}-m) \Psi \, , \\
M_{2NN} &=& i \Psib (4 c_1 m_{0\pi}^2) \Psi \, , \\
M_{3NN} &=& i \Psib (2 m_{0\pi}^4 (8 e_{38}+e_{115}+e_{116})) \Psi \, , \\
M_{4NN} &=& -\frac{ 3 i g_A^2}{4 \F^2} \Psib I_{\pi N}[p,\sla{\ell} 
\gamma_5 (\sla{p}+\sla{\ell}+m)\sla{\ell} \gamma_5)] \Psi \, , \\
M_{5NN} &=&-\frac{ 3 i  m_{0\pi}^2}{\F^2} \Psib  
(2 c_1-c_3-\frac{c_2 p^2}{d m_N^2}) \Psi I_\pi[1] \, , \\
M_{6NN} &=&-\frac{ 3 i  g_A^2}{4 \F^2}(-4 c_1 m_{0\pi}^2) \Psib 
I_{\pi NN}[p,p,\sla{\ell}\gamma_5 (\sla{p}+\sla{\ell}+m)
(\sla{p}+\sla{\ell}+m) \sla{\ell}\gamma_5] \Psi \, .
\eeqa

We evaluate these amplitudes using standard dimensional regularization
and expand the integrals in powers of the small momentum, keeping for
now the finite terms which do not obey counting. As discussed above
those finite terms which would be dropped in the Becher-Leutwyler
procedure are flagged with the symbol $\BL$. Likewise we flag terms
from each diagram which would be dropped in the EOMS procedure with
$\beta^{EOMS}$.

A first renormalization of the low energy constants is required to
ensure that the difference between the physical nucleon mass $m_N$ and
the nucleon mass in the chiral limit $\stackrel{\circ}{m_N}$ is
finite. In particular we take
\beqa
c_1^r&=&c_1 +\frac{3 g_A^2 m_N R}{128 \FPsq}(1-12 c_1 m_N)\, , \\
e_{115}^r+e_{116}^r+8e_{38}^r&=&e_{115}+e_{116}+8e_{38}+
\frac{3 R}{128 \FPsq}(c_2-8c_1+4 c_3- 168 c_1^2 m_N g_A^2) \, .
\eeqa

The finite renormalizations required read
\beqa
c_1^{IR} &=& c_1^{EOMS}\, , \\
e_{115}^{IR}+e_{116}^{IR}+8e_{38}^{IR}&=&e_{115}^{EOMS}+e_{116}^{EOMS}+
8e_{38}^{EOMS}-
\frac{3 {c_1}^2 m_N g_A^2}{16 \FPsq}(28+15 {\rm ln}(m_N^2/\mu^2)) 
\nn \\ & & -
\frac{3 g_A^2}{128 \FPsq m_N}(3+{\rm ln}(m_N^2/\mu^2)) \, , 
\eeqa
and
\beqa
c_1^{EOMS}&=&c_1^r-\frac{3 m_N g_A^2}{128 \FPsq}(1+8 c_1 m_N -
(1-12 c_1 m_N){\rm ln}(m_N^2/\mu^2))\, , \\
e_{115}^{EOMS}+e_{116}^{EOMS}+e_{38}^{EOMS}&=&e_{115}^r+e_{116}^r+
8e_{38}^r-
\frac{3 {c_1}^2 m_N g_A^2}{8 \FPsq}(8+3 {\rm ln}(m_N^2/\mu^2)) \, .
\eeqa

This leads to 
\beqa
m_N &=& \left[m-\frac{3 g_A^2 m^3}{32 \FPsq}(R+{\rm ln}(m^2/\mu^2))\right] - 
4 m_\pi^2 c_1^{IR}-2 m_\pi^4 (e_{115}^{IR}+e_{116}^{IR}+8e_{38}^{IR}) \nn \\
& & -
\frac{3 g_A^2 m_\pi^3}{32 \Fpi^2 \pi}-
\frac{3 g_A^2 m_\pi^4}{64 \FPsq m_N}(1+{\rm ln}(m_\pi^2/\mu^2))+
\frac{3 c_2 m_\pi^4}{128 \FPsq} \nn \\& & -
\frac{ m_\pi^4}{64 \FPsq}(3 c_2-32 c_1+12 c_3) {\rm ln}(m_\pi^2/\mu^2)  +
\frac{8 c_1 m_\pi^4}{\Fpi^2}l_3^r(\mu) \, .
\eeqa

Observe that all of the terms in the above formula which are
proportional to $1/\Fpi^2$ do not contribute to any physical amplitude
we calculate here. They would lead to $1/\Fpi^4$ corrections to loop
diagrams, the same order as two loop contributions which we are
neglecting. Furthermore the tree level diagrams do not contain nucleon
masses for which these terms are relevant. Thus for practical purposes
we can take
\beq
m \rightarrow m_N+4 c_1 m_\pi^2 +2 m_\pi^4(e_{115}+e_{116}+8e_{38}) \, ,
\eeq
where we have dropped all terms with $1/\Fpi^2$, which also 
allows $\{c_1^{IR},e_{115}^{IR},e_{116}^{IR},e_{38}^{IR}\} \rightarrow 
\{c_1,e_{115},e_{116},e_{38}\}  $.

Note also that in the chiral limit only the $R$ term survives,
i.e. we obtain 
\beq 
\stackrel{\circ}{m_N}= m-\frac{3 g_A^2 m^3}{32\FPsq}
(R+{\rm ln}(m^2/\mu^2)) \, .
\eeq 
It is common to introduce explicitly or implicitly a counterterm in
the Lagrangian to eliminate the correction term in this formula and
thus to interpret $m$ from the beginning as the nucleon mass in the
chiral limit, $\stackrel{\circ}{m_N}$, rather than as a 'bare' mass as
we have done.

Finally the nucleon wave function renormalization becomes
\beqa
\label{eq:ZN}
Z_N &=& 1-\frac{3m_\pi^2 g_A^2}{64 \FPsq}\left(2(1+\BL)+(3-2\BL)R+
16 c_1 m_N \BL(\EO-1)\right. \nn \\ & & + \left.
3{\rm ln}(\frac{m_\pi^2}{\mu^2})  -2 \BL(1-6 c_1 m_N (1-\EO))\LogmN-
\frac{3 m_\pi \pi}{m_N}\right)
\eeqa

If we reexpress the formulas for $m_N$ and $Z_N$ in terms of $m$ and
$m_{0 \pi}$ and take $\mu=m$ and take $\BL=0$ then these results agree
with those of Ref.~\cite{BL}. Note however that there are no LEC's or
counterterms available to absorb the $\BL$ or $\EO$ terms in $Z_N$,
i.e. those terms which are to be dropped in the IR or EOMS
schemes. Instead what happens is that these terms enter the amplitudes
via the $\sqrt{Z_N}$ terms which appear in
Eqs.~(\ref{eq:M1V}),(\ref{eq:M1A}),(\ref{eq:M2V}),(\ref{eq:M1pi}) and
are absorbed in LEC's elsewhere in the calculation.

\pagebreak

\begin{figure}
\begin{center}
\epsfig{figure=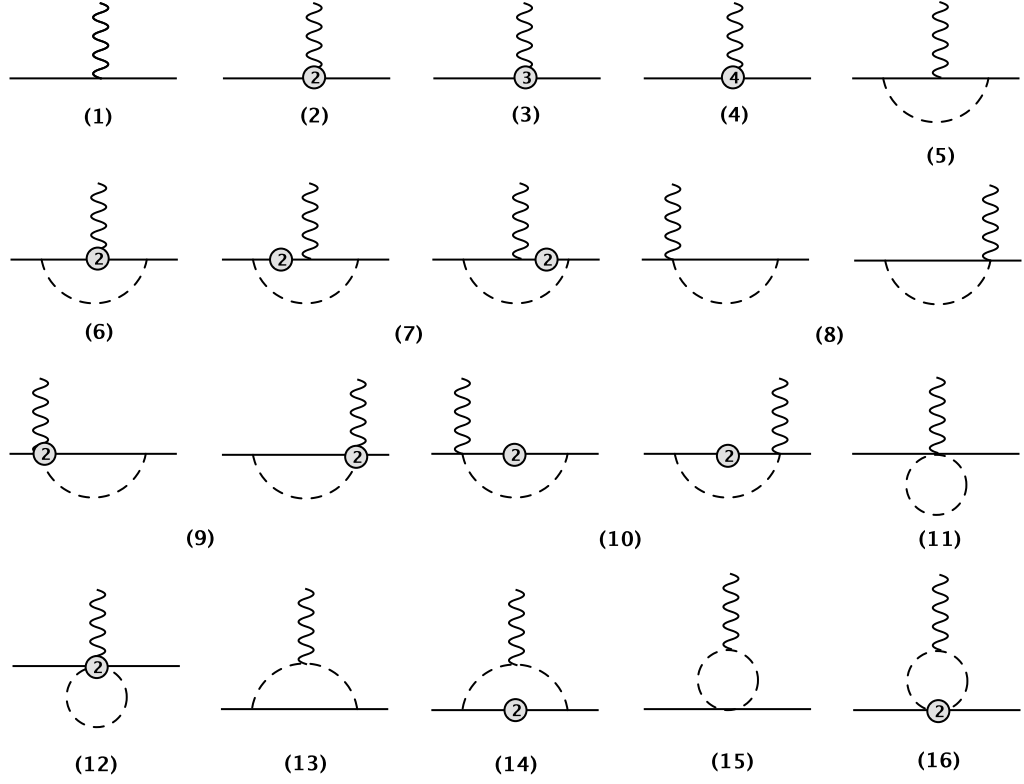, width=13.5cm,angle=0}
\caption{Diagrams which contribute to the coupling of external vector
and axial vector currents to the nucleon. The solid, dashed, and
wiggly lines correspond respectively to nucleons, pions, and external
vector or axial vector fields. The unlabeled vertices come from ${\cal
L}_{\pi N}^{(1)}$ whereas the ones labeled 2,3,4 come respectively
from ${\cal L}_{\pi N}^{(2)}$, ${\cal L}_{\pi N}^{(3)}$ and ${\cal
L}_{\pi N}^{(4)}$.}
\label{fig:avdiags}
\end{center}
\end{figure}

\begin{figure}
\begin{center}
\epsfig{figure=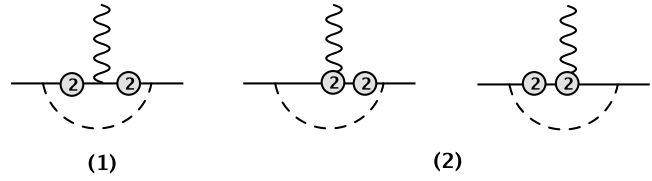, width=8.5cm,angle=0}
\caption{Diagrams which are higher order and which would not be
included in an explicit calculation. However parts of these diagrams
would be included implicitly by using the physical mass $m_N$ in the
propagators of loop integrals.  }
\label{fig:highord}
\end{center}
\end{figure}

\begin{figure}
\begin{center}
\epsfig{figure=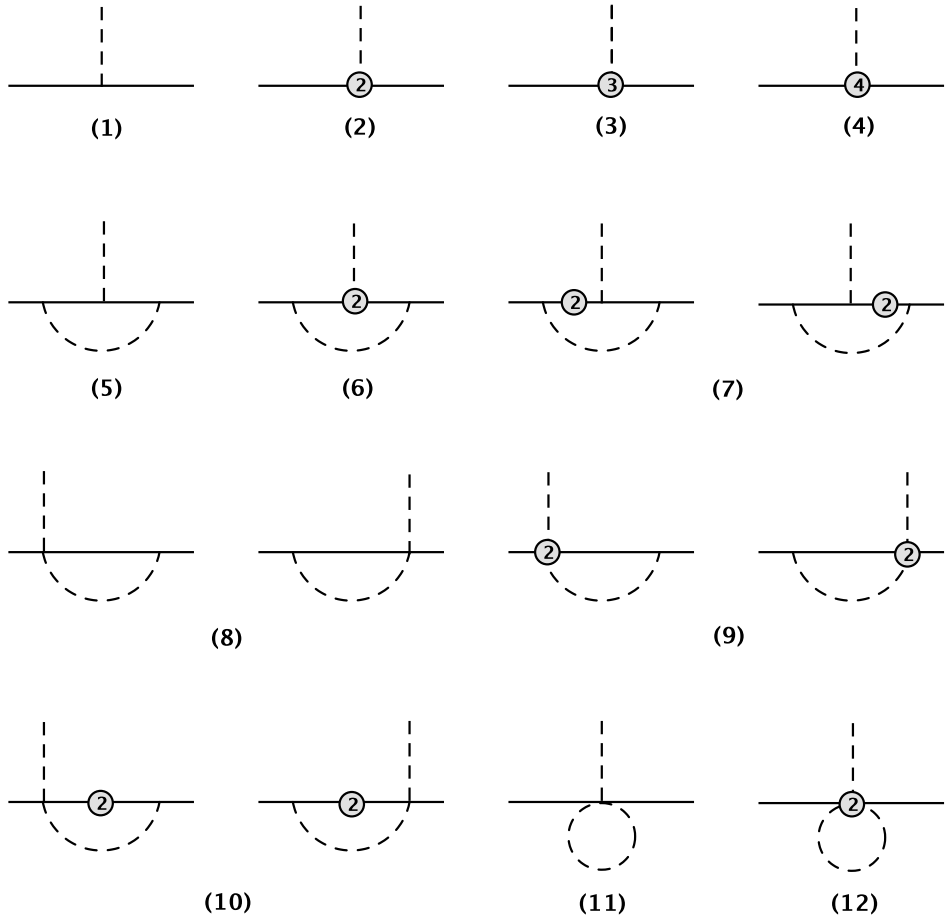, width=12.5cm,angle=0}
\caption{Diagrams which contribute to the $\pi$NN vertex. }
\label{fig:pnndiags}
\end{center}
\end{figure}

\begin{figure}
\begin{center} 
\epsfig{figure=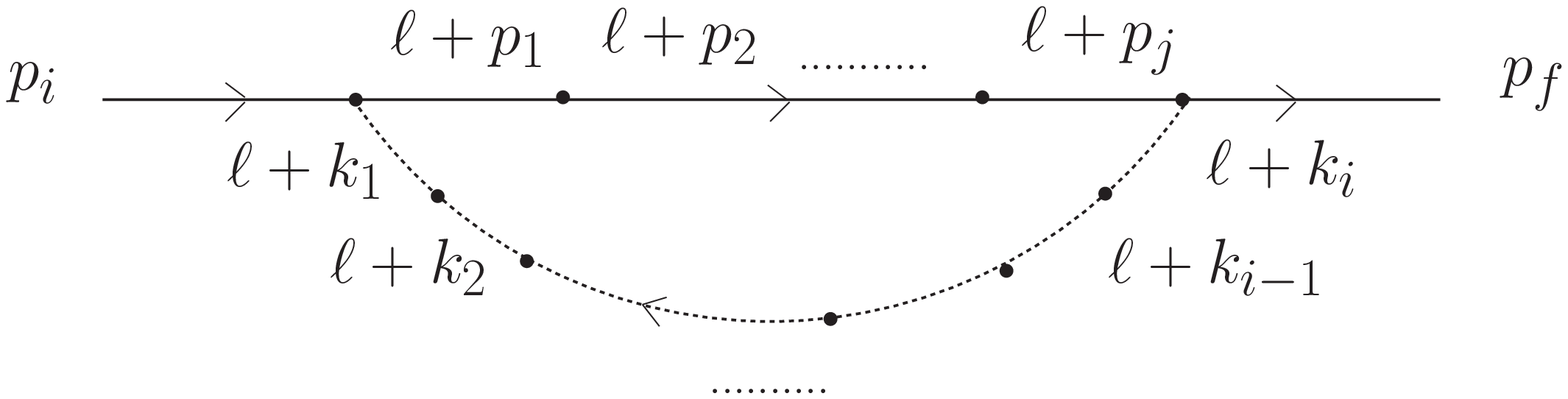, width=10.5cm,angle=0}
\caption{The general loop integral}
\label{fig:genloop}
\end{center}
\end{figure}

\begin{figure}
\begin{center}
\epsfig{figure=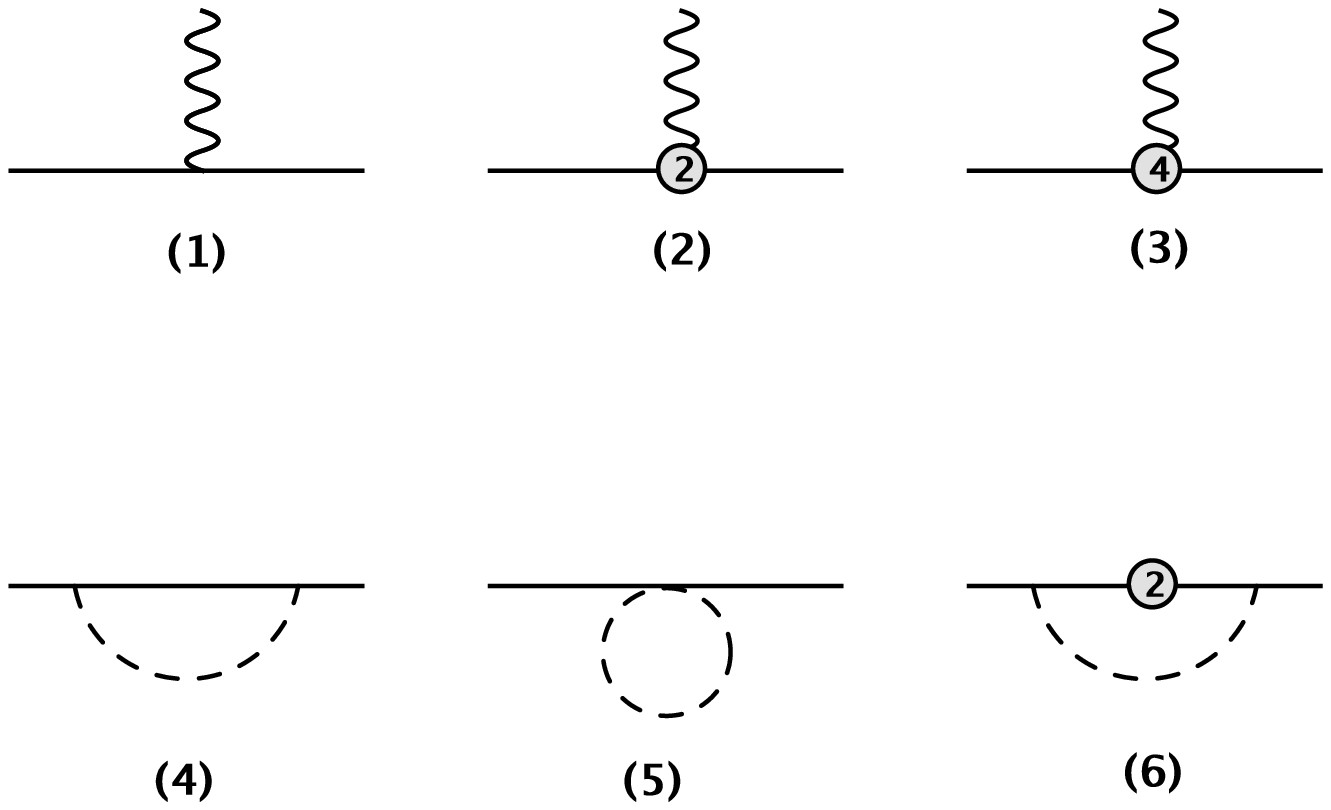, width=7.5cm,angle=0}
\caption{Diagrams contributing to the nucleon self energy}
\label{fig:nucselfen}
\end{center}
\end{figure}


\begin{thebibliography}{aaaaaaaaaaaaaaaaaaaaaaaaaaaaaa,syyyyn}
\bibitem{footnote1} 
Strictly speaking these are manifestly Lorentz
invariant theories rather than fully relativistic. They use
relativistic propagators and the full machinery of Dirac algebra, but
they do not include antiparticles as explicit degrees of freedom and
in particular do not include nucleon or anti nucleon loops. We will
refer to these theories somewhat imprecisely as 'relativistic' or
'manifestly Lorentz invariant' ChPT or as 'relativistic baryon ChPT',
or simply 'baryon ChPT' (BChPT).
\bibitem{footnote2}
It should be always be clear
from context that the integration variable $\ell$ is different from
the current $\ell_\mu$ defined earlier.
\bibitem{footnote3}
To obtain $I_{\pi \pi N}[-q,p_f,1]$ or $I_{\pi \pi
NN}[-q,p_f,p_f,1]$, which are not actually needed for this
calculation, one must take $p_i \rightarrow p_f$ in these formulas,
but also $x-1 \rightarrow x+1$ in $C_1$ and $z_0$. This latter change
results because $p_i,p_f,q$ are not all independent parameters, but
are related via $q=p_f-p_i$.
\end{thebibliography}
\end{document}